\pdfoutput=1 
\documentclass[a4paper,twoside]{article}
\newcommand{\affil}[1]{$^{\rm #1}$}
\usepackage[authoryear]{natbib}
\bibpunct{(}{)}{;}{a}{}{,}
\usepackage{graphicx}
\usepackage[fleqn]{amsmath}
\date{} 

\usepackage{amssymb}
\usepackage{colortbl} 
\usepackage{color} 
\definecolor{lightgray}{gray}{0.8}

\usepackage[pdftex, colorlinks=true, linkcolor=blue, citecolor=blue, filecolor=blue, urlcolor=blue, pdftitle=, pdfauthor=, pdfsubject=, pdfkeywords=,hyperfootnotes=false]{hyperref}

\title{\large\bf\flushleft Searching for Fast Radio Transients with SKA Phase 1}
\author{\parbox{\textwidth}{\flushleft
\vspace{-0.5cm}
{\it T. M. Colegate\affil{A, B} and N. Clarke\affil{A}}\\
\vspace{0.4cm}
{\small \affil{A}\,International Centre for Radio Astronomy Research, Curtin University, GPO Box U1987, Perth WA 6845}\\
{\small \affil{B}\,Email: tim.colegate@icrar.org}}}
\begin{document}
\twocolumn[
\begin{changemargin}{.8cm}{.5cm}
\begin{minipage}{.9\textwidth}
\vspace{-1cm}
\maketitle
%
%
\small{\bf Abstract:}

The Square Kilometre Array~(SKA) provides an excellent opportunity for low cost searches for fast radio transients. The increased sensitivity and field of view of the SKA compared with other radio telescopes will make it an ideal instrument to search for impulsive emission from high energy density events. 
We present a high-level search `use case' and propose \textit{event rate per unit cost} as a figure of merit to compare transient survey strategies for radio telescope arrays; we use \textit{event rate per beam formed and searched} as a first-order approximation of this measure.
Key results are that incoherent (phase insensitive) combination of antenna signals achieves the highest event rate per beam, and that 50--100 MHz processed bandwidth is sufficient for extragalactic searches with SKA Phase 1; the gain in event rate from using the full available bandwidth is small. Greater system flexibility will enable more effective searches, but need not drive the top-level system requirements beyond those already proposed for the SKA. The most appropriate search strategy depends on the observed sky direction and the source population; for SKA Phase~1, low frequency aperture arrays tend to be more effective for extragalactic searches and dishes more effective for directions of increased scatter broadening, such as near the Galactic plane.

\medskip{\bf Keywords:} surveys---telescopes---methods: observational---techniques: interferometric

\medskip
\medskip
\end{minipage}
\end{changemargin}
]
\small

\section{Introduction}

\begin{table*}[t!]
\begin{center}
\caption{Radio searches of the high time resolution universe and a comparison of event rate per beam.\label{tab:Experiments}}
\begin{tabular}{>{\raggedright}p{0.30\textwidth}>{\centering}p{0.13\textwidth}>{\centering}p{0.05\textwidth}>{\centering}p{0.05\textwidth}>{\centering}p{0.08\textwidth}>{\centering}p{0.12\textwidth}>{\centering}p{0.1\textwidth}}
\hline 
Experiment\textsuperscript{a} & Telescope and Status & $\nu_{\rm centre}$ (MHz) & $\Delta\nu$ (MHz) & Max. baseline (km)\textsuperscript{b} & $\mathcal{R}_{\rm beam^{-1}}$ (normalised)\textsuperscript{c} & Max. beams available \tabularnewline
\hline
Archival searches\textsuperscript{d} & Parkes & N/A & - & - & - & - \tabularnewline
Fly's eye fast radio transient search\textsuperscript{e} & ATA \\ (completed) & 1420 & 210 & N/A & $10^{-3}$ (fly) & 42 \tabularnewline
High Time Resolution Universe Pulsar Survey\textsuperscript{f} & Parkes \\ (operational) & 1352 & 340 & N/A & $10^{-2}$  & 13\tabularnewline
Pulsar ALFA (PALFA) Survey\textsuperscript{g} & Arecibo  \\ (operational)  & 1440 & 100 & N/A & $10^{-2}$ & 7\tabularnewline
V-FASTR\textsuperscript{h} & VLBA \\ (operational) & 1400 & 64 & 6000 & $10^{-2}$ (inc.) & 1\tabularnewline
LOFAR Transients Key Science Project\textsuperscript{i} & LOFAR \\ (in progress) & 120 & 32  & $<100$ & $10^{-1}$ (inc.) \\ $10^{-4}$ (coh.) &  1$^\dagger$ \\ thousands$^\dagger$ \tabularnewline
Commensal Real-Time ASKAP Fast-Transients (CRAFT) Survey\textsuperscript{j} & ASKAP \\ (planned) & 1400 & 300 & 6 & $10^{-2}$ (inc.) \\ $10^{-6}$ (coh.) & 36 \\ N/A\tabularnewline
Effelsberg Northern Sky Pulsar Survey\textsuperscript{k} & Effelsberg \\ (planned) & N/A & - & - & - & - \tabularnewline
SKA1 AA-low & & 260 & 380 & 200 & 1 (inc.) \\ $10^{-1}$ (coh.) & hundreds$^\dagger$ \\ thousands$^\dagger$ \tabularnewline
SKA1 low band dishes &  & 725 & 550 & 200 & 1 (inc.) \\ $10^{-2}$ (coh.) & 1 \\ thousands$^\dagger$\tabularnewline
\hline
\end{tabular}
\end{center}
\medskip
\textsuperscript{a}Only experiments within SKA1 frequencies (70 MHz -- 3 GHz) are listed. Pulsar surveys insensitive to single pulses are excluded. N/A is not applicable or information not available.\\
\textsuperscript{b}For event localisation using triggered buffer. \\
\textsuperscript{c}Order of magnitude estimation as per \autoref{eq:Event-rate-BW}, normalised to the incoherent combination of SKA1 low band dishes. For radio telescope arrays, the calculation is for fly's eye (fly), incoherent combination (inc.) or coherent combination (coh.), see \autoref{sec:Use-case}. A flat spectrum and no scatter broadening is assumed.\\
\textsuperscript{d}\citet{LorBai07, BurBai10, KeaKra11}.\\
\textsuperscript{e}\citet{SieBow11}.\\
\textsuperscript{f}\citet{KeiJam10}.\\
\textsuperscript{g}\citet{CorFre06, DenCor09}.\\
\textsuperscript{h}\citet{WayBri11}.\\
\textsuperscript{i}Dutch LOFAR as in \citet{HesSta09, LeeSta10}. More scenarios are discussed in \citet{StaHes11}.\\
\textsuperscript{j}\citet*{MacHal10}; \citet{MacBai10}.\\
\textsuperscript{k}\url{https://imprs-docs.mpifr-bonn.mpg.de/?p=45}.\\
$^\dagger$Limited by available beamformer processing and data transport.\\

\end{table*}

The high time resolution (HTR) universe has been relatively poorly observed at radio wavelengths and opening up this parameter space is a key science driver of the Square Kilometre Array~(SKA), via `Exploration of the Unknown' \citep{WilKel04}.
We define fast transients as impulsive, singly occurring or intermittent signals, emitted from high energy density events; a search for such events assumes an observed pulse width less than the normal correlator averaging time of a few seconds. In this context, pulsars can be classed as periodic fast transients. Other known transients include giant pulses, magnetars and rotating radio transients \citep{Cor07, MacBai10}. 

Interest in exploring HTR parameter space is growing; \autoref{tab:Experiments} lists some searches with existing and future telescopes. The SKA will provide at least an order of magnitude improvement in sensitivity over all these telescopes and a larger field of view (FoV) than most, as shown in Figure 1 of \citet{MacBai10}. 
Both FoV and sensitivity contribute to the expected rate of event detection; employing the analyses in \citet*{ CorLaz04} and \citet{Mac11}, this yields at least one or two orders of magnitude improvement in event rate for the SKA. 
\autoref{tab:Experiments} compares Phase 1 of the SKA (SKA1) to other experiments, using the calculations developed in this paper. 

Different search strategies are needed for fast transients and pulsars, despite some common HTR requirements. Pulsar surveys proposed for the SKA (e.g.~\citealp{SmiKra09, Cor07}) take advantage of pulsar periodicity to improve sensitivity. They involve a computationally expensive, systematic survey of the Galaxy, where the observer `drives' the telescope. The figure of merit (FoM) typically used to determine the effectiveness of such a survey measures the speed at which an \textit{area} of sky is surveyed to a certain sensitivity \citep{Cor07}. However, this FoM does not consider the number of events detectable in a \textit{volume} of sky, nor does it give any weight to the processing cost of sampling the sky or searching the data.

The main goal of at least first-generation fast transients searches is to maximise the number of events detected in a survey. 
The expected rate of event detection depends on, amongst other factors, the search strategy employed on the telescope. 
Given each search strategy has a different processing cost, \textit{event rate per unit cost} $(\mathcal{R}_{\rm cost^{-1}})$ is a more comprehensive FoM than survey speed.  Put simply, one search strategy may have a higher total rate of detection than another, but the processing cost of the strategy also needs to be considered. This is especially important for SKA1, considering transients detection will only be carried out `if it can be done with minimum additional cost or effort' \citep{Dewbij10}.

Because signal and search processing costs are architecture specific, we use a new FoM, \textit{event rate per beam formed and searched} ($\mathcal{R}_{\rm beam^{-1}}$) to generalise the problem and parametrise the effectiveness of a search strategy. 
It is based on the rate of transient events detectable in a \textit{volume} of sky as discussed in \citet{Mac11}, although it could similarly be applied to a FoM for surveying sky area. 
It assumes $\mathcal{R}_{\rm cost^{-1}}\propto\mathcal{R}_{\rm beam^{-1}}$, which is valid when cost increases linearly with the number of beams (independent FoVs) formed and searched. This is true for first-order beamforming and data transport costs for the SKA \citep*{ChiCol07, FauAle10}. The search costs also increase linearly because each beam signal needs to be searched individually;  we consider the efficiency gain from using a single processing unit to process multiple beams to be a second-order effect.

For a given search strategy, $\mathcal{R}_{\rm beam^{-1}}$ parametrises the choice of receptor (antenna), the performance, cost and efficiency of the signal combination mode and transients search system, and the observed sky direction. 
Because the cost of data storage is prohibitive\footnote{For example, $\sim20\,\rm GB$ per station beam per second would  need to be written to storage for SKA1 AA-low  ($\rm 380\,MHz\,bandwidth\times 2\,Nyquist\times4\,bits \times 50\, stations$).}, searches are conducted in real-time and data from candidate events recorded for subsequent verification and analysis.
To reduce processing costs, search strategies alternate to those proposed for pulsar surveys can be considered.  A pulsar survey need only visit each volume of sky once. In contrast, one volume of sky is considered as likely as another to contain transients \citep{Cor07}, and each time a volume is re-visited, there is new detection potential. This independence from sky direction enables lower cost commensal surveys for fast transients (piggy-back surveys where fast transients are not the primary telescope observation).

Understanding the most effective way to combine the signals from the antennas is also important. A radio telescope array produces images by correlating the antenna signals and averaging the output over a few seconds to reduce subsequent processing costs. To obtain a time resolution of order milliseconds or higher for fast transients detection, alternatives to `fast imaging' are currently required and various signal combination modes must be considered.  There are trade-offs between a highly sensitive mode with small FoV (such as the coherent combination of antenna signals) and less sensitive modes which cover more of the sky (incoherent combination, subarraying and `fly's eye'). These trade-offs are also influenced by the spatial density of antennas in the array. Furthermore, additional processing capability enables multiple beams to be formed and searched, re-using the array collecting area to some extent. The event rate per beam also depends on observing frequency and bandwidth. 

The SKA is being designed and constructed in two phases, SKA1 and SKA2, where the first is a subset of the second \citep{GarCor10}. This paper sets out a high-level `use case' for searching for fast transients with SKA1 receptors: low frequency aperture arrays and low band dishes (\autoref{sec:Use-case}). It outlines a basis for comparing survey strategies (\autoref{sec:Survey-strategy}) and undertakes a detailed analysis of the effects of receptor choice, signal combination modes, sky direction,  observing frequency and bandwidth on the event rate per beam (Sections \ref{sec:Trade-offs} and \ref{sec:Discussion}). These effects are summarised (\autoref{sec:Conclusions}) and specific recommendations for SKA1 are made (\autoref{sec:Recommendations}). This paper extends the work done by the International Centre for Radio Astronomy Research (ICRAR) for the Commensal Real-Time ASKAP Fast-Transients (CRAFT) survey \citep{MacHal10, MacBai10}. CRAFT is one of the survey science projects planned for the Australian SKA Pathfinder~(ASKAP),  a designated SKA precursor instrument. The analysis is applied to the SKA1 system description to determine optimal search strategies, but the method is equally applicable to SKA2 and other radio telescope arrays.

\begin{figure*}[t]
\begin{center}
\includegraphics[width=1.0\textwidth]{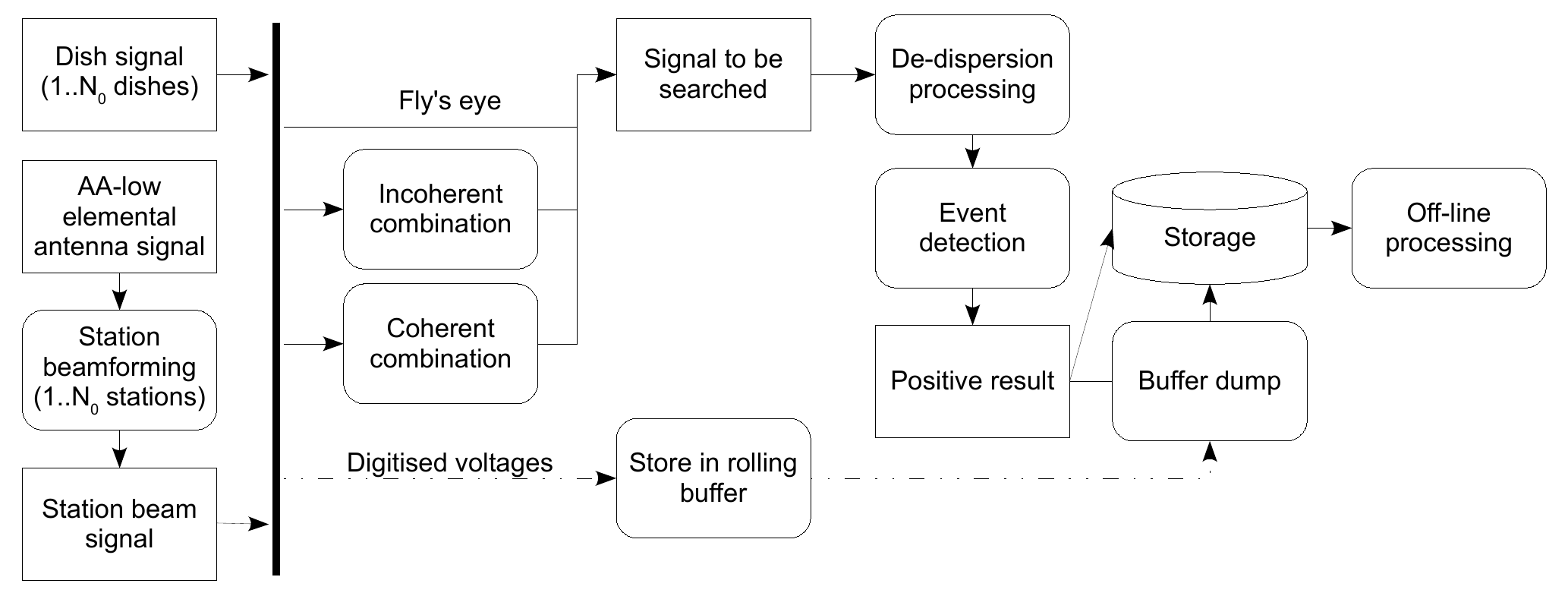}
\caption{High-level flow diagram for a generic fast transients pipeline, for SKA1 receptors. Rounded boxes are signal processing actions, rectangles describe the information flow. The solid vertical line is the data spigot point in the signal chain (see text for details).\label{fig:Pipeline}}
\end{center}
\end{figure*}

\begin{figure*}[t]
\begin{center}
\includegraphics[width=1.0\textwidth]{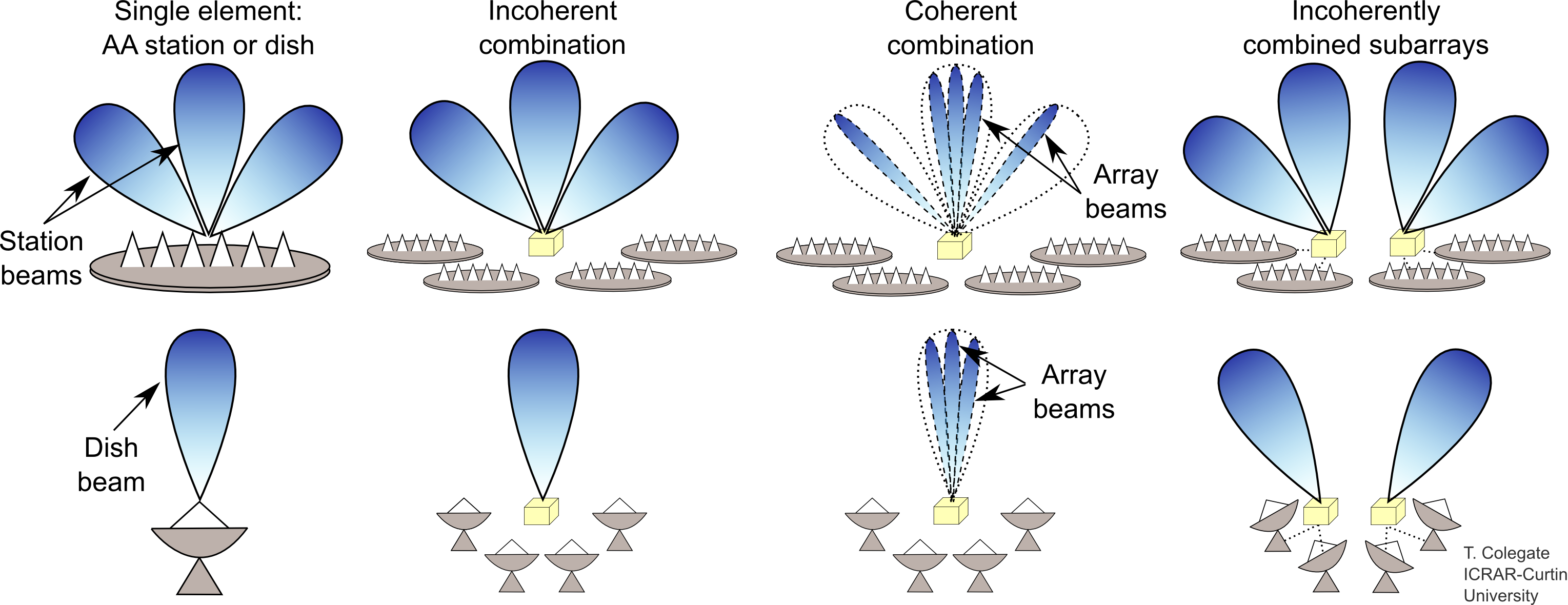}
\caption{Signal combination modes, resultant beam patterns, and beam terminology for dishes and aperture array (AA) stations (beam sizes not to scale).\label{fig:Signal-combination}}
\end{center}
\end{figure*}

\section{Fast transients search use case\label{sec:Use-case}}
For context, this section outlines a generic fast transients search use case. Because the data volumes are too large to store cost-effectively, the search for transient events is conducted in real-time on a data stream which is a continuous observation of the sky. However, a rolling buffer recording a small period of the data as it is observed allows candidate events 
containing potential fast transients detections to be saved and further processed off-line. The specific implementation of a fast transients search depends on the target or expected source population and the cost and performance factors of components in the processing pipeline; see \citet{MacBai10}, \citet{StaHes11} and \citet{WayBri11}. \autoref{fig:Pipeline} shows a generic fast transients pipeline and the signal processing steps (actions) in this pipeline are as follows:

\paragraph{Signal reception:}
Radio signals are collected by SKA1 receptors: low frequency (70--450~MHz) aperture arrays (AA-low) and a dish array equipped with low (0.45--1 GHz) and high (1--2~GHz) band single pixel feeds. Only the low band feed on the dishes (hereafter referred to as low band dishes)  is discussed in this paper, because it generally achieves a higher event rate than the high band feed, given the assumptions in \autoref{sec:Assumptions}.

\paragraph{Station beamforming:}
The complex signals of many AA-low elemental antennas are combined vectorially to form one or more station beams, as shown in \autoref{fig:Signal-combination}. The FoV of each station beam is determined by the station diameter and is similar to a single dish of the same diameter. The output station beam can then be processed in the same way as a dish signal. Though not considered here, signals from a group of dishes (instead of elemental antennas) can similarly be coherently combined into dish station beams. Note that both dish and AA stations are simply subarrays of coherently combined antennas.

\paragraph{Signal combination:}
The signals from the dish or station beams can be combined coherently, incoherently or not at all. These signals are then searched for fast transients. \autoref{fig:Signal-combination}  shows the signal combination modes and resultant beam patterns; they are further discussed in \autoref{sec:Signal-combination}. Incoherent (phase insensitive) combination sums the detected signals (powers) from antennas pointing in the same direction. Coherent combination of antennas forms a phased or tied array beam---voltages measured at each antenna are aligned in phase towards a specific direction on the sky, in a manner similar to station beamforming. Smaller groups of antennas---subarrays---can be incoherently combined and each subarray pointed in a different direction. The extreme of this is the so-called fly's eye, where every antenna is pointed in a different direction. 

Event localisation and the spatial discrimination of astronomical signals from radio frequency interference (RFI) is possible for coherent combination and, using buffered voltages, for incoherent combination and subarrays of three or more incoherently combined antennas. Multiple beams (incoherently or coherently combined) can be used to discriminate RFI, where a candidate event in most or all beams indicates the presence of RFI.

\paragraph{Dedispersion processing:}
The signals pass through a cosmic medium of unknown dispersion measure (DM). This means that the detection needs to be trialled for many DMs, each of which has a computational cost. The DM range to be trialled depends on the location on the sky. \citet{ClaDAd11} discusses dedispersion for SKA1 in detail.

\paragraph{Event detection:}
An event detection algorithm needs to be applied to the signal from each trial DM, where optimal detection is achieved with an appropriate matched filter \citep{CorMcl03}.

\paragraph{Store in rolling buffer:}
The digitised voltages from the dishes or stations are stored in a circular memory (rolling) buffer. In the case of a candidate event, the data from the buffer can be saved to another location (dumped) and processed off-line. 
The amount of memory required in the buffer depends on the sampling rate, sample size and the expected maximum (dispersed) pulse duration. The maximum pulse duration is a function of the range of frequencies to be captured and varies linearly with the maximum DM to be trialled. For a maximum DM of $\rm 3000\,pc\,cm^{-3}$ and a bandwidth of a few hundred MHz, a buffer of order tens of seconds is required for dish frequencies and possibly tens of minutes for lower frequencies.

\paragraph{Buffer dump and off-line processing:}
On receipt of a trigger, the buffer will dump the original voltage data to storage for off-line processing, which could include RFI filtering, analysis of the candidate detection and correlation of the dish or station beams for source localisation and imaging.

\paragraph{Commensal and targeted surveys:}
A commensal survey greatly increases observation time by conducting the survey in parallel with normal telescope operations. It is passive;  it uses dish or station beam signals from the primary user observation, placing little extra demand on the telescope. Such a survey is suitable for extragalactic searches, given the information about the population of such fast transients is not known a priori; hence one direction on the sky is as good as another. To observe specific areas of the sky, such as the Galactic plane and nearby galaxies, a targeted transients survey (which is the primary user observation) may be required (e.g.~\citealp{LeeSta10}).

\paragraph{Data spigot}
A data spigot to the dish and station beam signals is useful for transients surveys, especially those which are commensal.
If the signal chain is considered to be the signal path from the antennas of a radio telescope array to the correlator, a spigot defines a point in the signal chain where users can tap off data via a well defined interface.  The spigot for fast transient searches may output either coherent (phase-preserved) data at high rates or, alternatively, incoherent data where the dish or station beam voltages are squared and integrated to a time resolution of order milliseconds to reduce the data rate and subsequent dedispersion processing load.  The latter approach is being taken by CRAFT to access beams from the ASKAP beamformer \citep{MacHal10}. Similarly, searching the integrated signals from the dish or station beams which have been incoherently combined is a low cost option for commensal surveys with SKA1.

The solid vertical line in \autoref{fig:Pipeline} shows the point in the flow diagram where the spigot for fast transients would need to exist to enable the signal combination modes in this paper. The pipeline after the spigot point is not part of the normal imaging mode of the telescope; the post-spigot pipeline may be implemented internally or with user-provided processing. 
An example of processing being implemented internally is the `non-imaging processing' for pulsar observations with SKA1 \citep{Dewbij10}; the approach being taken by CRAFT is an example of user-provided processing. Note that it is conceivable that a spigot to the AA-low elemental antenna signals could also exist, but the data rates make this option prohibitively expensive for SKA1.

\section{Survey strategy: Maximising survey speed and minimising cost\label{sec:Survey-strategy}}

Figures of merit quantify the effect of altering the variable parameters of a problem. A simple FoM to measure the cost effectiveness of a fast transients search strategy is the detected event rate per beam searched ($\mathcal{R}_{\rm beam^{-1}}$), a proxy for cost in the absence of sufficiently accurate design and cost information. 
We want to optimise for a high value of $\mathcal{R}_{\rm beam^{-1}}$, although the total event rate for all beams ($\mathcal{R}_{\nu}$) must also be high enough to be of scientific benefit and open new volumes of parameter space. The qualitative advantages of a strategy, which cannot be captured in a FoM, must also be considered. 

The detected event rate is effectively a survey speed. \citet{SmiKra09} present a frequency-dependent FoM for survey speed (SSFoM) for dishes. It is based on surveying an \textit{area} of sky, thus SSFoM is linearly proportional to FoV and sensitivity squared. In this paper,  the equivalent SSFoM is the rate of transient events detectable in a \textit{volume} of sky and is linearly proportional to FoV  and sensitivity to the power of~3/2. It draws on event rate calculations from \citet{Mac11}; the derivation is shown in \autoref{sec:Rate-derivation}.

The event rate is given by
\begin{equation}\mathcal{R_{\nu}} = \rho\frac{\Omega_{\rm proc}}{4\pi}V_{\rm max}\quad {\rm events\,s}^{-1},\label{eq:Event-rate-volume}\end{equation}
where $V_{\rm max}$ is the maximum volume out to which an object is detectable and $\Omega_{\rm proc}$ is the processed FoV, which is the product of the number of beams formed ($N_{\rm beam}$) and the FoV of each beam.
Thus the event rate per beam is
\begin{equation}
\mathcal{R}_{\rm beam^{-1}}=\frac{\mathcal{R}_{\nu}}{N_{\rm beam}}.
\end{equation}
$N_{\rm beam}$ may be the number of station ($N_{\rm b-0}$) or array ($N_{\rm b-arr}$) beams. For subarrays, it is the product of the number of subarrays and station beams formed in each subarray ($N_{\rm sa}N_{b-0}$) .

We describe an extragalactic survey as a search for a homogeneously distributed population of isotropically emitting fast transients of fixed intrinsic luminosity. For such a population, the event rate is\begin{equation}
\mathcal{R}_{\nu}=\frac{1}{3}\rho\Omega_{\rm proc}\left(\frac{W_{\rm i}}{W}\right)^{\frac{3}{4}}\left(\frac{\mathcal{L}_{\rm i}}{4\pi S_{{\rm min}}}\right)^{\frac{3}{2}}\quad {\rm events\,s}^{-1},\label{eq:Event-rate}\end{equation}
 where $\rho\ ({\rm events\,s^{-1}\,pc^{-3}})$ is the event rate density, $\mathcal{L}_{\rm i}\,({\rm Jy\, pc^{2}})$ is the intrinsic luminosity of the population, $W_{\rm i}$ is the intrinsic pulse width, $W$ is the observed pulse width and $S_{{\rm min}}$ is the minimum detectable flux density of the telescope for an integration time of $\tau=W_{\rm i}$. The $(W_{\rm i}/W)^{3/4}$ term approximates the loss in signal-to-noise (S/N) due to pulse broadening (see \autoref{sec:Rate-derivation}). The  frequency dependence of  $\mathcal{R}_{\nu}$ is discussed in more detail in \autoref{sub:Event-rate-vs-freq}.

\citet{Mac11} shows that the event rate is proportional to $\Omega_{\rm proc}$ and $S_{\rm min}^{3/2}$ for an extragalactic population with a luminosity distribution which follows a power-law or lognormal distribution. Although the actual event rate depends on the luminosity distribution, the proportionality still holds for $\Omega_{\rm proc}$ and $S_{\rm min}$, which is sufficient to compare telescopes and their signal combination modes. 
For fast transients searches within the Galaxy, scatter broadening due to multipath propagation in the interstellar medium makes the event rate per beam dependent on frequency and direction. Although comprehensive direction-dependent modelling is beyond the scope of this paper,  the loss in sensitivity due to scattering is incorporated in the $W$ term in \autoref{eq:Event-rate} and modelled for some representative sky directions. This first-order analysis gives an indication of how SKA1 will perform as a function of frequency; see \autoref{sec:Galactic} for further discussion of Galactic objects.

\section{Modelling event rates\label{sec:Trade-offs}}

Event rate per beam is a simple metric to model fast transients event rates in the cost constrained environment of SKA1. This section shows how signal combination mode and filling factor (antenna spatial density) affect $\mathcal{R}_{\rm beam^{-1}}$ for a radio telescope array. The dependence on frequency and the effectiveness of searching large bandwidths are also considered. The modelling is specifically applied to the SKA1 receptors, for source populations whose intrinsic luminosity either does not vary with frequency or varies with $\nu^{-1.6}$, a value typical of the pulsar population \citep{LorYat95}. This paper uses the system description of  SKA1 \citep{Dewbij10} to make trade-offs---the relevant details are given in \autoref{tab:SKA1}. However, the complete system design for SKA1 is still under development and subject to a decision-making process involving trade-offs and performance and cost optimisation. 

\begin{table}[h!]
\begin{center}
\caption{SKA1 system details.\label{tab:SKA1}}
\begin{tabular}{>{\raggedright}p{0.55\columnwidth}>{\raggedleft}p{0.4\columnwidth}}
\hline 
\multicolumn{2}{c}{\textbf{Low frequency aperture arrays (AA-low)}} \tabularnewline
\textbf{Aperture} &  \tabularnewline
Frequency range\textsuperscript{a} & 70--450 MHz \tabularnewline
Station diameter ($D_{0}$) & 180 m \tabularnewline
Number of stations ($N_{0}$) &50 \tabularnewline
Number of antennas ($N_{\rm a}$) & 11 200 per station \tabularnewline
Station beam taper ($\mathcal{K}_{0}$) & 1.3 \tabularnewline
Dense--sparse transition ($\nu_{\rm transition}$) & 115 MHz (2.6 m)  \tabularnewline
\multicolumn{2}{l}{\textbf{Array configuration regions}\textsuperscript{b} } \tabularnewline
Core (radius$<$0.5 km)  & $\sim$ 50\% (25 stations)  \tabularnewline
Inner (1$<$radius$<$2.5 km)  & $\sim$ 20\% (10 stations)  \tabularnewline
Mid (2.5$<$radius$<$100 km)  & $\sim$ 30\% (15 stations)  \tabularnewline
Core filling factor  & 0.81\tabularnewline
\textbf{Performance} &  \tabularnewline
T$_{\rm rcvr}$  & 150 K \tabularnewline
Bandwidth per beam ($\Delta\nu$) & 380 MHz\tabularnewline
\hline 
\multicolumn{2}{c}{\textbf{Single pixel feed dishes}} \tabularnewline
\textbf{Aperture}  &  \tabularnewline
SKA2 dish frequency capability & 0.3--10 GHz \tabularnewline
Parabolic dish diameter ($D_{0})$ & 15 m  \tabularnewline
Number of dishes ($N_{0}$) & 250 \tabularnewline
Total physical aperture  & 44 179 m$^{2}$ \tabularnewline
Dish illumination factor ($\mathcal{K}_{0}$) & 1.15 \tabularnewline
\multicolumn{2}{l}{\textbf{Array configuration regions}\textsuperscript{b} } \tabularnewline
Core (radius$<$0.5 km)  & $\sim$ 50\% (125 ant.) \tabularnewline
Inner (0.5$<$radius$<$2.5 km)  & $\sim$ 20\% (50 ant.)  \tabularnewline
Mid (2.5$<$radius$<$100 km)  & $\sim$ 30\% (75 ant.) \tabularnewline
Core filling factor  & 0.03 \tabularnewline
\textbf{Antenna RF system}\textsuperscript{c} &  \tabularnewline
Feed/LNA low band & 0.45--1.0 GHz \tabularnewline
Bandwidth ($\Delta\nu_{\rm low}$)  & 0.55 GHz  \tabularnewline
Feed/LNA high band & 1.0--2.0 GHz \tabularnewline
Bandwidth ($\Delta\nu_{\rm high}$)  & 1.0 GHz \tabularnewline
\textbf{Performance} &   \tabularnewline
Antenna/feed efficiency\textsuperscript{d}  & 70\% \tabularnewline
Average $T_{\rm sys}$ in low band\textsuperscript{e} & $\sim$40 K \tabularnewline
Average $T_{\rm sys}$ in high band & $\sim$30 K \tabularnewline
\hline
\end{tabular}
\end{center}
\medskip
\textsuperscript{a} Single dual polarization antenna over frequency range. \\
\textsuperscript{b} Fractional number in each region.\\
\textsuperscript{c} One dual polarization feed available at a time.\\
\textsuperscript{d} Average over frequency.\\
\textsuperscript{e} Higher at the low frequency end of this band. \\ \\ \\ \\ \\
\end{table}

\subsection{Assumptions\label{sec:Assumptions}}
The trade-offs in this section make the following simplifying assumptions:
\begin{itemize}
\item The population of fast transients is homogeneously spatially distributed and of fixed intrinsic luminosity.
\item A matched filter is used to detect the dedispersed, but scatter broadened pulse (as per \citealp{CorMcl03}).
\item The effects of scintillation on source intermittency and optimum search bandwidth are ignored.
\item The dedispersion processing system does not contribute to pulse broadening (see \autoref{sec:Broadening} for a description of these instrumental contributions).
\item The intrinsic pulse width is 1 ms. Shorter duration pulses would be more sensitive to S/N loss due to pulse broadening, longer duration pulses less sensitive.
\item Events are broad-band such that the intrinsic spectral bandwidth of the pulse is greater than the processed bandwidth. Thus all channels across the band contain contributing signal.
\item The beam has constant (maximum) sensitivity between the half-power beamwidth points, and zero sensitivity outside of that. 
\item Beamformer calibration costs are not considered.
\item A time to frequency domain transformation (channelisation) and cross-correlation `FX' correlator is used, as opposed to other correlator topologies such as `XF'. It is the most cost-effective architecture for the SKA, and allows other signal processing actions, such as beamforming and RFI excision, to be done efficiently \citep{HalSch08}.
\item The processing cost of forming and searching a beam is independent of frequency, bandwidth and signal combination modes. In practice, lower frequencies (where the maximum dispersed pulse duration is longer) and larger bandwidths will increase processing costs; the magnitude of the increase is specific to the processing architecture and the effect of sky direction on the DM range to be trialled. 
\end{itemize} 

\subsection{Signal combination mode comparisons\label{sec:Rate-modes}}
This section compares the per beam event rate for incoherent and coherent combination and fly's eye (\autoref{fig:Signal-combination} and \autoref{sec:Signal-combination}) and applies these results to SKA1 AA-low and low band dishes. We refer to a single dish or AA station as an element,  designated with the subscript 0. The combination of an array of $N_{0}$ elements may refer to the number of elements in the total array, or some subset of the total array (e.g. the SKA1 core region in \autoref{tab:SKA1}).

Incoherent combination of an array of $N_{0}$ elements increases the sensitivity by a factor of $\sqrt{N_{0}}$ over a single element while retaining its FoV, $\Omega_{0}$. Forming $N_{\rm b-0}$ station beams linearly increases the FoV. To even further increase the FoV, $N_{\rm sa}$ subarrays can be incoherently combined. Each subarray is pointed in a different direction, increasing the FoV by a factor of $N_{\rm sa}$ but only increasing the sensitivity of the array by a factor of $\sqrt{N_{\rm 0/sa}}$ over a single element, where $N_{\rm 0/sa}$ is the number of elements per subarray. Fly's eye pertains to the case $N_{\rm 0/sa}=1$.

Sensitivity of the coherent combination of an array of $N_{0}$ elements is higher than incoherent combination and subarraying; it increases proportional to $N_{0}$. However the FoV of the array beam, $\Omega_{\rm arr}$, is much smaller; it is proportional to $D_{\rm arr}^{-2}$, where $D_{\rm arr}$ is the diameter of the array of elements being combined. The FoV can be linearly increased by forming $N_{\rm b-arr}$ array beams.

Applying these relationships to \autoref{eq:Event-rate} gives the total event rate for each signal combination mode:\begin{align}
\begin{split} 
\mathcal{R}_{\nu}  = &  \frac{1}{3}\rho\left(\frac{W_{\rm i}N_{\rm pol}\Delta\nu\tau}{W}\right)^{\frac{3}{4}}\left(\frac{\mathcal{L}_{\rm i}A_{\rm e0}}{4\pi\sigma2k_{\rm B}T_{\rm sys}}\right)^{\frac{3}{2}}\mathcal{M}, \\
\mathcal{M} =  & \begin{cases}
N_{\rm b-0}\Omega_{0}N_{0}^{3/4} & \rm{Incoherent\,combination}\\
N_{\rm b-arr}\Omega_{\rm arr}N_{0}^{3/2} & \rm{Coherent\,combination}\\
N_{\rm sa}^{1/4}N_{\rm b-0}\Omega_{0}N_{0}^{3/4} & \rm{Subarraying},\end{cases}
\end{split}\label{eq:Event-rate-modes}\end{align}
where $N_{\rm pol}$ is the number of polarisations summed, $\Delta\nu$ is the processed bandwidth, $\tau$ is the post-detection integration time (which also defines the time resolution of the observation), $A_{\rm e0}$ is the effective area of an element (dish or station), $\sigma$ is the S/N ratio required for event detection and $T_{\rm sys}$ is the system temperature.

\subsubsection{Filling factor efficiency}
The coherent combination mode is more effective if the dishes or stations are closely spaced, thus having a higher filling factor. In this case, the same number of elements are being combined, but the array beam FoV is larger. \citet{DAd10} considers the number of coherently combined array beams required to achieve an event rate FoM equivalent to one incoherently combined beam for ASKAP. We modify this analysis and apply it to $\mathcal{R_{\nu}}$. 

Following \citet{Cor07}, we define the number of pixels ($N_{\rm pix}$) as the maximum number of independently pointed, coherently combined array beams which can be formed within the FoV of a single element beam. It is frequency independent, and given by
\begin{align}
\begin{split}
N_{\rm pix} & = \frac{\Omega_{0}}{\Omega_{\rm arr}}\\
 & = \left(\frac{\mathcal{K}_{0}}{\mathcal{K}_{\rm arr}}\frac{D_{\rm arr}}{D_{0}}\right)^{2}\label{eq:N_pix},\end{split}\end{align}
where $\mathcal{K}_{0}$ and $\mathcal{K}_{\rm arr}$ are the element and array beam tapers respectively.

A measure of the effectiveness of the coherent combination mode is the number of array beams which need to be formed and searched to achieve a coherent combination event rate ($\mathcal{R}_{\rm coh}$) equal to incoherent combination ($\mathcal{R}_{\rm inc}$). From \autoref{eq:Event-rate-modes}, $\mathcal{R}_{\rm coh}=\mathcal{R}_{\rm inc}$ gives
\begin{align}\begin{split}
N_{\rm b-arr} & = \frac{N_{\rm b-0}\Omega_{0}N_{0}^{3/4}}{\Omega_{\rm arr}(\eta N_{0})^{3/2}}\\
 & = \frac{N_{\rm b-0}N_{\rm pix}}{\eta^{3/2} N_{0}^{3/4}},\label{eq:Num-array-beams}
\end{split} \end{align}
where $\eta$ is the fraction elements in the array which are coherently combined, out of a total $N_{0}$. 
For example, $\eta=0.5$ if only the elements in the SKA core are coherently combined while those in the total array are incoherently combined.

Achieving the highest possible event rate is desirable, but this must be tempered by the cost of searching multiple beams. The relative event rate per beam depends on the array filling factor and is simply the inverse of \autoref{eq:Num-array-beams} when $N_{\rm b-0}=1$:
\begin{equation}
\mathcal{R}_{\rm coh\,beam^{-1}} =  \frac{\eta^{3/2}N_{0}^{3/4}}{N_{\rm pix}}\mathcal{R}_{\rm inc\,beam^{-1}}.\label{eq:Relative-rate-coherent}
 \end{equation}
 For incoherently combined subarrays (when all $N_{0}$ elements are formed into subarrays),
\begin{equation}
\mathcal{R}_{\rm sa\,beam^{-1}} = N_{\rm sa}^{-3/4}\mathcal{R}_{\rm inc\,beam^{-1}},\label{eq:Relative-rate-subarray}\end{equation}
where $N_{\rm sa}$ subarray beams are searched. 

\subsubsection{Coherent combination for a fully filled array\label{sub:Ideal-case}}
Assuming a best case scenario of an array entirely filled with stations of equal diameter and $\mathcal{K}_{0}=\mathcal{K}_{\rm arr}$,
\begin{equation}
N_{\rm pix} = N_{0}.
\end{equation}
Substituting into \autoref{eq:Num-array-beams} and for $\eta=1$, the theoretical minimum number of array beams required so that $\mathcal{R}_{\rm coh}=\mathcal{R}_{\rm inc}$ is
\begin{equation}N_{\rm b-arr} = N_{\rm b-0}N_{0}^{1/4}\end{equation} 
and the relative event rate per beam (\autoref{eq:Relative-rate-coherent}) is
\begin{equation}
\mathcal{R}_{\rm coh\,beam^{-1}} = N_{0}^{-1/4}\,\mathcal{R}_{\rm inc\,beam^{-1}}.
\end{equation}
This estimation is optimistic towards coherent combination, given that it is not physically possible to entirely fill a circular array with circular stations.
Regardless of this, for $N_{0}>1$, \textit{the incoherent combination will always achieve a higher event rate per beam searched than coherent combination, and this difference increases with $N_{0}$.} These results are frequency independent.

\begin{table*}[t]
\begin{center}
\caption{Relative event rates for SKA1 receptors and select signal combination modes\label{tab:Event-rate}}
\begin{tabular}
{>{\columncolor{lightgray}\raggedright}p{0.25\textwidth}>{\columncolor{lightgray}\centering}p{0.1\textwidth}>{\columncolor{lightgray}\centering}p{0.1\textwidth}>{\centering}p{0.1\textwidth}>{\centering}p{0.15\textwidth}>{\centering}p{0.15\textwidth}}
\hline
Signal combination mode & \multicolumn{2}{>{\cellcolor{lightgray}}c}{Input parameters\textsuperscript{a}} &  \multicolumn{3}{c}{Calculated values} \tabularnewline
& $D_{\rm arr}$\,(km) & $N_{0}$ & $N_{\rm pix}$ & $N_{\rm b-arr}$ required\textsuperscript{b} & Relative $\mathcal{R}$ per beam\textsuperscript{c}\tabularnewline
\hline
\multicolumn{3}{>{\cellcolor{lightgray}}c}{AA-low receptors} &&&\tabularnewline
Incoherent: total array & - & 50 & - & - & 1\tabularnewline
Coherent: core & 1 & 25 & 31 &  $5N_{\rm b-0}$ & $2.15\times10^{-1}$\tabularnewline
Coherent: inner + core & 5 & 35 & 772 & $70N_{\rm b-0}$  & $1.43\times10^{-2}$\tabularnewline
Coherent: total array  & 200 & 50 & $1.23\times10^{6}$ & $6.57\times10^{4}N_{\rm b-0}$  & 
$1.52\times10^{-5}$\tabularnewline
Fly's eye: total array & - & 50 subarrays & - & - & $5.32\times10^{-2}$\tabularnewline
\hline
\multicolumn{3}{>{\cellcolor{lightgray}}c}{Low band dish receptors} &&&\tabularnewline
Incoherent: total array & - & 250 & - & - & 1\tabularnewline
Coherent: core & 1 & 125 & $3.49\times10^{3}$ & 157 & $6.37\times10^{-3}$\tabularnewline
Coherent: inner + core & 5 & 175 & $8.72\times10^{4}$ & $2.37\times10^{3}$ & $4.22\times10^{-4}$\tabularnewline
Coherent: total array  & 200 & 250 & $1.40\times10^{8}$ & $2.22\times10^{6}$ & $4.50\times10^{-7}$\tabularnewline
Fly's eye: total array & - & 250 subarrays & - & - & $1.59\times10^{-2}$\tabularnewline
\hline
\end{tabular}
\end{center}
\textsuperscript{a} From \citet{Dewbij10}.\\
\textsuperscript{b} Number of array beams required for an event rate equivalent to incoherent combination. For dishes, $N_{\rm b-0}=1$.\\
\textsuperscript{c} Relative to the incoherent combination event rate per beam.\\
\end{table*}

\subsubsection{Signal combination modes for SKA1}
A flexible processing system allows various signal combination modes; \autoref{tab:Event-rate} compares the event rate for incoherent and coherent combination and fly's eye, for SKA1 receptors. The number of array beams required such that $\mathcal{R}_{\rm coh}=\mathcal{R}_{\rm inc}$ is calculated for coherent combination. The event rate per beam formed and searched, relative to $\mathcal{R}_{\rm inc\,beam^{-1}}$, is calculated for all modes. Three coherent combination modes are shown: elements in the core region, inner and core region and the total array. Of these, using the elements in the core achieves the highest event rate per beam, due to the higher density of collecting area. For this reason, the core will be the only coherent combination mode further analysed in this paper.

The coherent combination of the AA-low core requires approximately five array beams to equal the event rate of a single beam of the incoherently combined total array.  Although the fly's eye mode achieves a higher event rate by a factor of $50^{1/4}=2.6$, the relative rate per beam is less than the coherently combined core.
If the dense packing is not achievable \citep{Dewbij10}, the number of array beams required will be higher. Indeed, using optimisations from \citet{GraLub98}, the optimal packing of 25 congruent circles of diameter 180 m in a circle results in a minimum core diameter of 1036 m. This is larger than the 1000 m diameter in \citet{Dewbij10} and excludes any spacing that may be required for infrastructure.

Due to the lower filling factor, 157 beams formed from the coherent combination of the dishes in the core are required to equal the event rate of the incoherently combined array. The event rate of the fly's eye mode is higher by a factor of $250^{1/4}=4$ and its relative rate per beam is higher than the coherently combined core. Although not shown, values for subarrays lie between fly's eye and incoherent combination modes, and depend on the number of elements per subarray.

\begin{figure*}[t]
\begin{center}
\includegraphics[height=0.26\textheight]{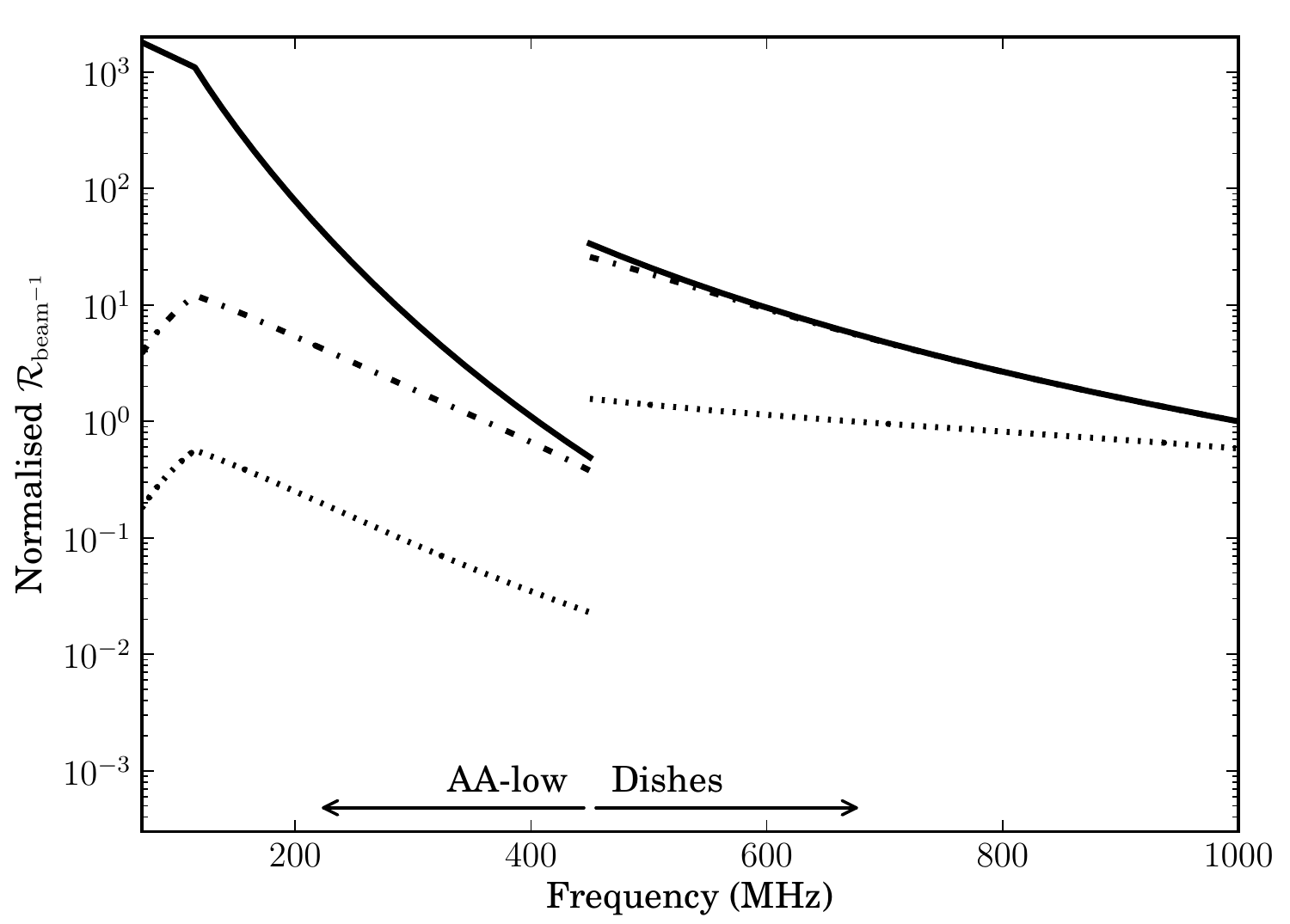}\includegraphics[height=0.26\textheight]{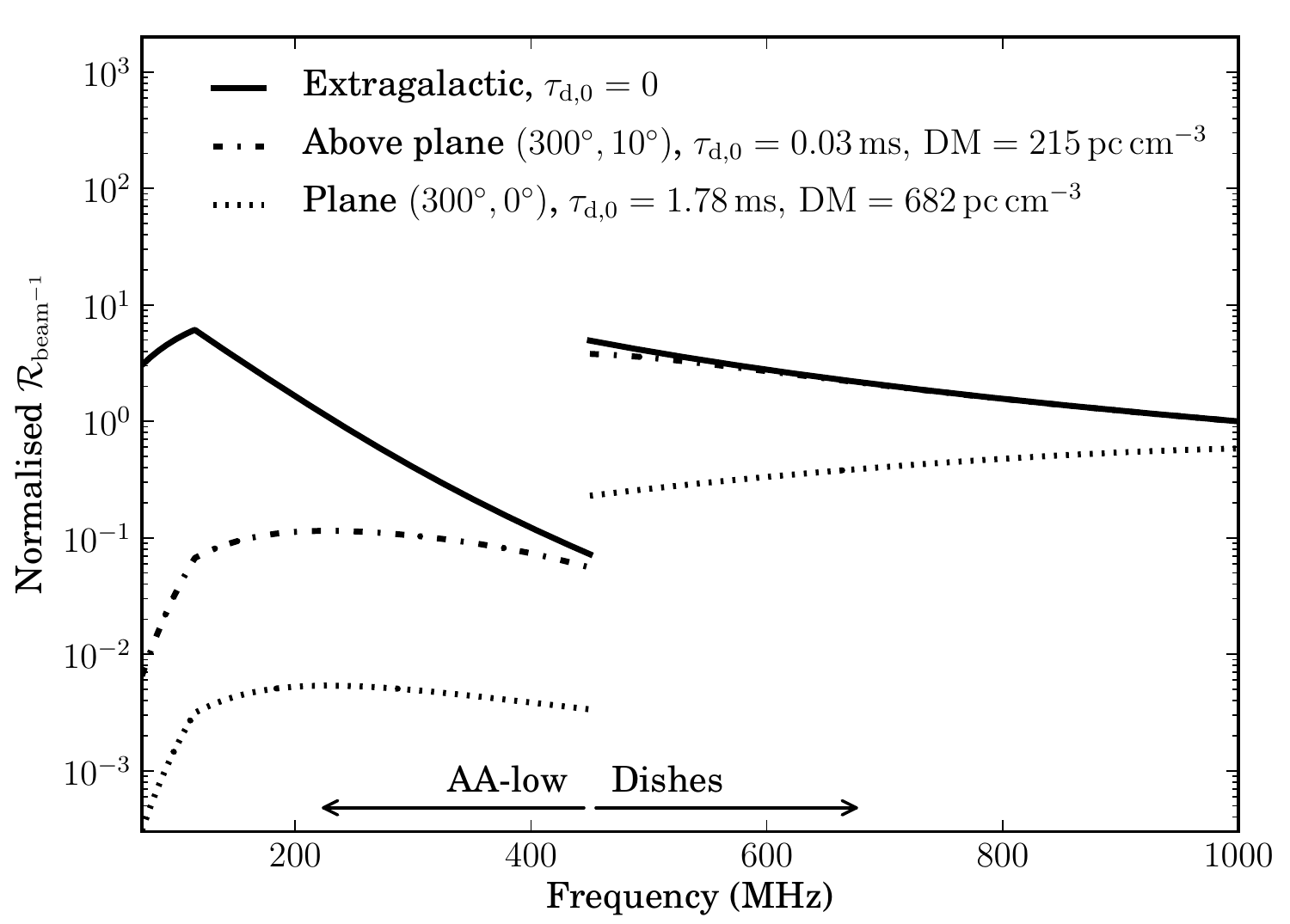}
\caption{Normalised event rate per beam for the incoherent combination of the total array for $\Delta\nu=1\,{\rm MHz}$ and a source spectral index of -1.6 (left) and 0 (right), at three representative sky directions. The rate for the coherent combination of the elements in the array core is less by a factor of approximately five for AA-low  and 157 for the low band dishes. The post-detection integration time equals the intrinsic pulse width: $\tau=W_{\rm i}=1\,{\rm ms}$. Scatter broadening and DM is calculated at $\nu_{0}=1\,{\rm GHz}$ from NE2001 \citep{CorLaz02} for a distance of 30 kpc and scaled using $\tau_{{\rm d}}\propto\nu^{-4.4}$. Data is normalised to $\mathcal{R}_{{\rm beam^{-1}}}=1$ at $\nu_{0}$ for the extragalactic case. The aperture array dense-sparse transition frequency is at 115~MHz. \label{fig:Event-rate-vs-freq}}
\end{center}
\end{figure*}

\subsection{Frequency dependence\label{sub:Event-rate-vs-freq}}

The event rate has a frequency dependence on luminosity, minimum detectable flux density, FoV and scatter broadening, designated by subscript $\nu$:
\begin{equation}
\mathcal{R}_{\nu} =  \frac{1}{3}\rho\Omega_{\rm proc_{\nu}}\left(\frac{W_{\rm i}}{W_{\nu}}\right)^{\frac{3}{4}}\left(\frac{\mathcal{L}_{\rm i,\nu}}{4\pi S_{{\rm min}_{\nu}}}\right)^{\frac{3}{2}}.\label{eq:Event-rate-fn-freq}
\end{equation}
It is important  to understand this frequency dependence because of the wide fractional bandwidths of the SKA. 
Looking at each of these dependencies in turn:
\begin{itemize}
\item The processed FoV depends on the number of beams formed and whether they are formed incoherently or coherently, but either way is proportional to $\nu^{-2}$:
\begin{equation}
\Omega_{\rm proc_{\nu}}=\frac{\pi}{4}N_{\rm beam}\left(\frac{{\rm c}\mathcal{K}}{\nu D}\right)^{2},
\end{equation}
where c is the speed of light.
For incoherent combination, $N_{\rm beam}$ is the number of station beams formed, $\mathcal{K}$ is the feed illumination factor or station beam taper and $D$ is the diameter of the dish or station. For coherent combination, $N_{\rm beam}$ is the number of array beams formed, $\mathcal{K}$ is the array beam taper and $D$ is the diameter of the array (see \autoref{sec:Signal-combination}).
\item Pulses are broadened due to scattering. The broadening time $\tau_{{\rm d}}$ depends on the path through the Galaxy to the observer, and scales as $\tau_{{\rm d}}\propto\nu^{-4.4}$ \citep{CorLaz02}.
The observed pulse duration is given by
\begin{equation}
W_{\nu}\approx\sqrt{W_{\rm i}^{2}+\tau_{d}^{2}}.
\end{equation}
See \autoref{sec:Broadening} for further details.
\item Because we are looking for an unknown population, we do not know how
the luminosity varies with frequency. However, spectral indices for
pulsars have been measured. We use $\xi=-1.6$ \citep{LorYat95}, a value typical of the pulsar population, such that\begin{equation}
\mathcal{L}_{\rm i,\nu}=\mathcal{L}_{0}\left(\frac{\nu}{\nu_{0}}\right)^{\xi},\end{equation}
where $\mathcal{L}_{0}$ is luminosity at reference frequency $\nu_{0}$.  For comparison, we also consider a flat spectrum population ($\xi=0$).
\item $S_{{\rm min}_{\nu}}$ is a function of $T_{\rm sys_{\nu}}$ and $A_{\rm e0_{\nu}}$:
\begin{equation}
S_{{\rm min}_{\nu}} \propto \frac{T_{\rm sys_{\nu}}}{A_{\rm e0_{\nu}}}.
 \end{equation}
For aperture arrays, the effective area of a station is approximately\begin{equation}
A_{\rm e0_{\nu}}=\begin{cases}
\frac{\pi}{4}D_{\rm 0}^{2} & \nu<\nu_{\rm transition}\\
N_{\rm a}\times\frac{c^{2}}{3\nu^{2}} & \nu>\nu_{\rm transition}.\end{cases}\end{equation}
The AA-low system temperature is the sum of the receiver noise and an approximation
to the sky temperature:\begin{equation}
T_{\rm sys_{\nu}}=T_{\rm rcvr}+60\left(\frac{c}{\nu}\right)^{2.55}.\end{equation}
\end{itemize}
A graphical breakdown of the frequency dependencies of AA-low is shown in \autoref{sub:AA-low-Appendix}.

The dependence of event rate on frequency for three representative sky directions and spectral indices of $\xi=0$ or $-1.6$ is shown in \autoref{fig:Event-rate-vs-freq}. The normalised $\mathcal{R}_{\rm beam^{-1}}$ is plotted at 1~MHz intervals for centre frequency $\nu$ and processed bandwidth $\Delta\nu=1\, {\rm MHz}$, spanning the SKA1 system description frequency range of AA-low and low band dishes (70 MHz $\leq\nu\leq$ 1000 MHz). 
The relative event rate per beam between signal combination modes (the rightmost column of \autoref{tab:Event-rate}) still applies as a multiplicative factor to the data in \autoref{fig:Event-rate-vs-freq}, regardless of sky direction and frequency.

The simplest case to consider is a search for extragalactic fast transients (solid line in \autoref{fig:Event-rate-vs-freq}). In this case there is no sensitivity loss due to scatter broadening; $W\gg\tau_{\rm d}$ is assumed. 
\citet{CorMcl03} find that for a given sky direction, the scatter broadening of an extragalactic source will be approximately six times the broadening from the Galaxy alone. This assumes equal scatter broadening in the host galaxy, if there is one, and no contribution from  the intergalactic medium. For directions away from the Galactic plane where $\tau_{\rm d}$ is low and for an intrinsic pulse width of 1 ms, the exclusion of scatter broadening is a reasonable first-order assumption. 

A first-order analysis of a search for Galactic transients is possible by invoking the simplifying assumptions listed in \autoref{sec:Assumptions}. We calculate relative event rates taking into account estimates of scatter broadening at a distance of 30~kpc and frequency of 1~GHz from the NE2001 model of \citet{CorLaz02}, where the broadening scales as $\tau_{{\rm d}}\propto\nu^{-4.4}$. A distance of 30~kpc determines the maximum broadening due to interstellar scattering for that direction; broadening is less at shorter distances. Two representative sky directions for Galactic transients are:
\begin{itemize}
\item above the Galactic plane: $\tau_{{\rm d}}=0.03\,\rm ms$ ($DM=215\,{\rm pc\, cm^{-3}}$) at l = 300, b = 10
\item on the Galactic plane: $\tau_{{\rm d}} =1.78\,\rm ms$ ($DM=628\,{\rm pc\, cm^{-3}}$) at l = 300, b = 0.
\end{itemize}
For these directions, the normalised $\mathcal{R}_{\rm beam^{-1}}$ in \autoref{fig:Event-rate-vs-freq} shows how increased scatter broadening reduces the event rate at lower frequencies. In this first-order analysis, the effect of scatter broadening on event rate is independent of the signal combination mode chosen; the frequency-dependent $(W_{\rm i}/W)^{3/4}$ term in \autoref{eq:Event-rate-modes} is common to all modes. The factors to consider for a more detailed analysis of Galactic transients searches is outlined in \autoref{sec:Galactic}.

From \autoref{fig:Event-rate-vs-freq},  the incoherent combination of dish signals would be most efficient for searching for extragalactic sources with a low spectral index. AA-low would be more efficient for such sources with high spectral indices. 
For a Galactic population, the preferred receptor depends on the amount of scatter broadening. For directions near or on the Galactic plane,  low band dishes show a higher $\mathcal{R}_{\rm beam^{-1}}$ than AA-low. However, the difference between the two receptors is less than an order of magnitude for steep spectrum sources near the Galactic plane.
For steep spectrum sources, even those on the Galactic plane, the higher event rate is at the lower end of the frequency band of each receptor. For shallow spectrum sources near or on the Galactic plane, there is no strong maximum within a receptor frequency band. For dishes, the maximum may be at the low (450 MHz) or high (1 GHz) end of the frequency band, for small or large $\tau_{\rm d}$ respectively.
For AA-low, the event rate quickly reduces for frequencies below $\nu_{{\rm transition}}=115\,\rm MHz$.

\begin{table*}[t]
\begin{center}
\caption{Normalised extragalactic event rate per beam\textsuperscript{a} for the full SKA1 receptor bandwidth. \label{tab:Event-rate-BW}}
\begin{tabular}{>{\raggedright}p{0.3\textwidth}>{\centering}p{0.15\textwidth}>{\centering}p{0.15\textwidth}>{\centering}p{0.15\textwidth}>{\centering}p{0.15\textwidth}}
\hline
Receptor  & \multicolumn{2}{c}{Incoherently combined: total array} &  \multicolumn{2}{c}{Coherently combined: core}  \tabularnewline
& $\xi$=-1.6 & $\xi$=0 & $\xi$=-1.6 & $\xi$=0 \tabularnewline
\hline
Low band dish ($\rm \Delta\nu = 550\,MHz$) & $989$ & $262$ & $6.30$ & $1.67$\tabularnewline
AA-low ($\rm \Delta\nu =  380\,MHz$) & $3.13\times10^{4}$ & $164$ & $6.74\times10^{3}$ & $35.3$\tabularnewline
AA-low and low band dish & $3.15\times10^{4}$ & $361$ & $6.74\times10^{3}$ & $35.7$\tabularnewline
\hline
\end{tabular}
\end{center}
\medskip
\textsuperscript{a} Normalised to $\mathcal{R}_{{\rm beam^{-1}}}=1$ ($\Delta\nu=1\,\rm MHz$) for the incoherent combination of array elements at $\nu_{0}=1\,{\rm GHz}$. The actual event rate per beam is $6.94\times 10^{-8}\rho\mathcal{L}_{0}^{3/2}x\,\rm events\,s^{-1}$, where $x$ is the normalised $\mathcal{R}_{{\rm beam^{-1}}}$, $\rho$ has units of $\rm events\,s^{-1}\,pc^{-3}$ and $\mathcal{L}_{0}$ has units of $\rm Jy\,pc^{2}$.
\end{table*}

\begin{figure}[t]
\begin{center}
\includegraphics[width=1\columnwidth]{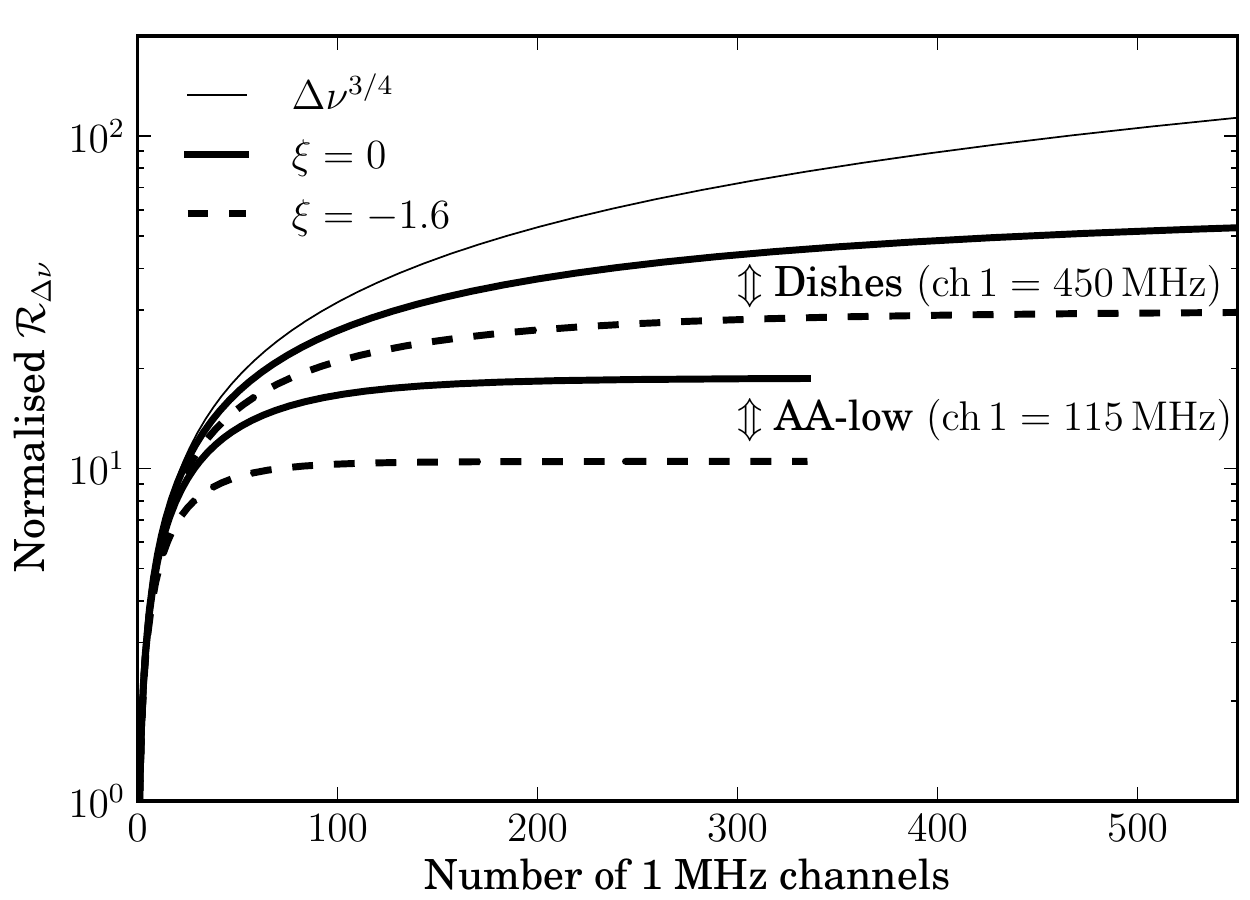}
\caption{Extragalactic event rate, calculated for 1~to~335 AA-low channels of width 1~MHz, where channel~1 is at 115~MHz, and 1~to~550 low band dish channels, where channel~1 is at 450~MHz.
Thick lines show the calculated $\mathcal{R}_{\Delta\nu}$ versus processed bandwidth $\Delta\nu=N_{\rm ch}\rm\,MHz$, for spectral indices of 0 and -1.6.  The thin line is the expected $\Delta\nu^ {3/4}$ increase in event rate over the rate using only one channel ($\mathcal{R}_{\Delta\nu=1\,{\rm MHz}}$).
The rate for each curve is normalised such that $\mathcal{R}_{\Delta\nu=1\,\rm MHz}=1$; the slope of each curve applies equally to the incoherent or coherent combination of the receptor.\label{fig:Event-rate-BW}}
\end{center}
\end{figure}

\subsection{Large processed bandwidths\label{sub:Event-rate-bandwidth}}
Processed bandwidth ($\Delta\nu$) is the bandwidth of the astronomical signal at the fast transients detection system. In radio astronomy, an increase in processed bandwidth is usually assumed to produce a $\sqrt{\Delta\nu}$ increase in signal-to-noise. From \autoref{eq:Event-rate-modes}, this would produce a $\Delta\nu^{3/4}$ increase in event rate. However, unless the event rate is approximately constant across frequency, this increase does not hold for large processed bandwidths.

To calculate the event rate over a large processed bandwidth, the frequency-dependent channel contributions shown in \autoref{fig:Event-rate-vs-freq} are summed.  Where the pulse broadening (due to propagation and instrumental effects) is not significant, the event rate is calculated numerically for a processed bandwidth of $\Delta\nu=N_{\rm ch}\Delta\nu_{\rm ch}$ using
 \begin{align}\begin{split}
\mathcal{R}_{\Delta v} & =\left(\sum_{i}^{N_{\rm ch}}\mathcal{R}_{\nu,i}^{4/3}\right)^{\frac{3}{4}}\\
& =\frac{1}{3}\rho\left(\frac{\mathcal{L}_{0}}{\nu_{0}^{\xi}}\right)^{\frac{3}{2}}\left(\sum_{i}^{N_{\rm ch}}\left(\frac{\Omega_{{\rm proc},i}^{2/3}\nu_{i}^{\xi}}{S_{{\rm min},i}}\right)^{2}\right)^{\frac{3}{4}},\end{split}\label{eq:Event-rate-BW}\end{align} 
where $N_{\rm ch}$ is the number of frequency channels of width $\Delta\nu_{\rm ch}$. The derivation is shown in \autoref{ap:App-Event-Rate}. \autoref{tab:Event-rate-BW} makes these calculations for different modes and spectral indices, for the full bandwidth of AA-low and low band dishes. The relative event rate per beam between signal combination modes of a given receptor (Equations \ref{eq:Relative-rate-coherent} and \ref{eq:Relative-rate-subarray}) still hold for large processed bandwidths. Strikingly, for the combined AA-low and low band dish event rate, the contribution from the dishes is only significant for the incoherently combined total array case, when $\xi=0$. 

Given $\Delta\nu=N_{\rm ch}\Delta\nu_{\rm ch}$, plotting $\mathcal{R}_{\Delta\nu}$ as a function of the number of contributing channels shows the decreasing contribution of higher frequency channels to the event rate.  This decreasing contribution results in the event rate increasing by less than $\Delta\nu^{3/4}$. The only exception is in the dense AA regime ($<115\,\rm MHz$) for low spectral indices ($-1\lesssim\xi\le0$), where the decreasing contribution comes from lower frequency channels.

Consider two cases for extragalactic searches, plotted in \autoref{fig:Event-rate-BW}: the AA-low band from 115 to 450 MHz and the low band dishes where the whole 550 MHz bandwidth is available. The ideal case of a $\Delta\nu^ {3/4}$ increase over the $\Delta\nu=1\,{\rm MHz}$ event rate is also shown. As expected, this plot shows that the maximum bandwidth achieves the highest event rate. However, the event rate curve flattens out well before the maximum bandwidth, especially for AA-low and also steeper spectrum sources.

Assuming a limited amount of signal processing is available, the following question arises: at what point can the processing be more effectively used elsewhere in the fast transients pipeline, and how is this quantified? One method is to arbitrarily set a threshold beyond which additional channels contribute very little to the event rate. Channels of increasing frequency are included while the following is true:
\begin{equation}
\frac{\mathcal{R}_{\Delta\nu+\Delta\nu_{\rm ch}}}{\mathcal{R}_{\Delta\nu}}>{\rm threshold}.
\end{equation}
For example, say the threshold is set to 0.5\%. Then for the ideal case, $N_{\rm ch}$ = 150 channels contribute to $\mathcal{R}_{\Delta\nu}$. Adding a 151$^\text{st}$ channel will contribute less than 0.5 \% to $\mathcal{R}_{\Delta\nu}$. For the two cases plotted, \autoref{tab:Contributing-channels} shows the maximum channel for which the improvement in event rate over the rate without that channel is greater than 0.5\%. 

\begin{table}[h]
\begin{center}
\caption{Maximum number of channels of bandwidth $\Delta\nu_{\rm ch}=\rm 1\,MHz$ contributing more than 0.5\% of the cumulative event rate. For the ideal case, the maximum channel number is 150.\label{tab:Contributing-channels} }
\begin{tabular}{>{\raggedright}p{0.35\columnwidth}>{\centering}p{0.25\columnwidth}>{\centering}p{0.25\columnwidth}}
\hline
Case  & \multicolumn{2}{c}{Channel number}  \tabularnewline
& $\xi=-1.6$ & $\xi=0$ \tabularnewline
\hline
AA-low,  $\rm ch\,1 = 115\,MHz$& 47 & 69\tabularnewline
Low band dishes, $\rm ch\,1 = 450\,MHz$ & 87 & 112\tabularnewline
\hline
\end{tabular}
\end{center}
\end{table}

To interpret this table, compare the maximum contributing channel number when $\xi=-1.6$.  For AA-low, 47 channels ($\Delta\nu=47\,{\rm MHz}$) contribute above the threshold. For dishes, this is achieved with $\Delta\nu=87\,{\rm MHz}$, implying that the AA-low processed bandwidth becomes less useful more quickly. This is expected, given the steeper spectral dependence of AA-low over dishes, shown in \autoref{fig:Event-rate-vs-freq}. 

\section{Discussion\label{sec:Discussion}}
\subsection{Effectiveness of combination modes}
Comparing signal combination modes and array configuration using event rate per beam ($\mathcal{R}_{\rm beam^{-1}}$) gives a frequency independent analysis of the trade-offs for a given receptor. Incoherent combination always achieves a higher $\mathcal{R}_{\rm beam^{-1}}$ than coherent combination, as shown in \autoref{sub:Ideal-case}. The difference increases with the number of elements or a reduced filling factor. The per beam analysis is important because it captures the first-order processing costs for each beam. For this reason, although fly's eye gives the highest total event rate, it comes at the expense of searching many more beams.

However, the per beam analysis does have some limitations. The total event rate $\mathcal{R}_{\nu}$ of a search strategy is
\begin{equation}
\mathcal{R}_{\nu}=\mathcal{R}_{\rm beam^{-1}}N_{\rm beam},
\end{equation}
where $N_{\rm beam}$ is the number of beams formed and searched. Depending on the signal combination mode used, $N_{\rm beam}$ may be the number of station ($N_{\rm b-0}$) or array ($N_{\rm b-arr}$) beams, or $N_{\rm sa}N_{\rm b-0}$ for $N_{\rm sa}$ incoherently combined subarrays. Electromagnetic and signal processing system design and cost considerations will put maxima on each of these. 
For example, the number of station beams which can be formed within the antenna FoV is limited by performance degradation, the diameter of the station and the beamforming processing power available. Once the limit on forming more station beams is reached,  $\mathcal{R}_{\nu}$ will eventually become higher for coherent rather than incoherent combination. This is because $N_{\rm pix}$ array beams can be formed within each of the $N_{\rm b-0}$ station beams (\autoref{eq:N_pix}), allowing a maximum of $N_{\rm pix}N_{\rm b-0}$ array beams to be formed.

Dishes with single pixel feeds only have a single beam (i.e.~$N_{\rm b-0}=1$), so $\mathcal{R}_{\nu}$ of the incoherent combination of dishes cannot be increased by forming more beams. However, if phased array feeds (PAFs) are available on dishes, then $N_{\rm b-0}$ dish beams can be formed, increasing the total event rate.

So we must consider: do we want to form and search more array beams, subarrays or incoherently combined station beams? This will be influenced by station beamforming (for AAs) and array beamforming costs, search costs and the array filling factor. Although exact costs are unavailable, we can generalise preferred combination as follows:
\begin{itemize}
\item \textbf{Incoherent combination}

Achieves the highest event rate per beam, making it preferable in most cases. If the cost of searching a beam signal is high, incoherent combination presents a further advantage over other modes.

\item \textbf{Coherent combination}

Requires an array with a high filling factor and low beam search costs; it can achieve the highest total event rate if many array beams are be formed and searched. For AA-low, the cost of forming multiple coherently combined array beams must be lower than the cost of forming and incoherently combining multiple station beams. 

\item \textbf{Incoherently combined subarrays and fly's eye}

This mode is only preferable when the beam search cost is low. A fly's eye mode excludes the buffering and source localisation advantages of an array, and commensality with most observations (see \autoref{sec:Use-case}). Sets of three-element subarrays counter this problem and could employ antennas unused by the primary user observation. For example, splitting 24 AA-low stations outside the core into 8 of these three-element  subarrays results in a total event rate approximately equal to the incoherent combination of 50 stations.
\end{itemize}

The SKA1 AA-low results show a weak preference for the incoherent combination of station beams: five array beams formed from the coherent combination of the AA-low stations in the core are required to equal the event rate of a single beam of the incoherently combined total array, assuming the stations are very closely packed. If multiple station beams are formed `at no cost' for normal (imaging) array observing, or the cost of array beamforming is high, the advantage of incoherent combination is increased. In practice, other effects such as RFI mitigation and station and array beam quality also need to be taken into account.

For SKA1 low band dishes, 157~array beams from the coherent combination of dishes in the core are required to equal the event rate of the incoherently combined total array. Coherent combination with dishes would only be optimal for hundreds of array beams, and even then it is likely that the processing power would be more effectively spent on AA-low. However, an incoherent commensal survey with low band dishes would be a cost-effective method to cover parameter space.

\subsection{Frequency and bandwidth effects}
The \textit{relative} event rate per beam between signal combination modes described in the previous section is independent of frequency for searches of extragalactic populations and the first-order analysis of Galactic populations. The frequency-dependent effects in \autoref{eq:Event-rate-modes} are common to all modes; the exception being processed FoV, however that cancels when the relative rate is calculated.

For a given signal combination mode, the event rate per beam changes with frequency, as shown in \autoref{fig:Event-rate-vs-freq}. For the extragalactic case (or more generally, when the observed pulse width is much greater than the scatter broadening), the slope on the dish event rate is due to the spectral index of the source and changing FoV; for aperture arrays, the frequency dependence of $T_{\rm sys}$ (due to sky noise) and $A_{\rm e}$ is also a factor. Additionally, some pulsars display a turnover (a break in the spectrum where pulsar brightness is maximum) around 100--200~MHz \citep{MalGil94}, and others above 1~GHz \citep{KijLew11}. The event rate for such sources would decrease below the turnover frequency. For the Galactic case, the scatter broadening increasingly reduces the S/N for lower frequencies, in addition to the other factors described above.

For SKA1, $\mathcal{R}_{\rm beam^{-1}}$ is generally higher for AA-low than dishes, however the opposite is true for sources with low spectral indices and directions of larger scatter broadening. Also, lower frequencies have an increased memory cost for dedispersion. When low spectral indices, increased scatter broadening or low-frequency turnovers are factors and only a few AA station or array beams are available, the incoherent combination of the low band dishes gives a higher $\mathcal{R}_{\nu}$.  If SKA1 dishes are equipped with PAFs, this would increase the total incoherent combination event rate of the dish array, as multiple station beams does for AA-low.

The frequency and bandwidth effects on event rate per beam are interdependent. \autoref{sub:Event-rate-bandwidth} shows that an optimal  frequency range for searching for fast transients with SKA1 may be smaller than the full band of the receptor. As the frequency increases, the FoV reduces and the source luminosity is expected to decrease for sources with pulsar-like emission characteristics. Processing more channels (hence bandwidth) increases $\mathcal{R}_{\nu}$, however this increase is less than $\Delta\nu^{3/4}$. 

For some threshold beyond which extra bandwidth contributes little to the event rate, the processing for channels above this threshold could be more effectively used to form and search extra beams (increasing $\mathcal{R}_{\nu}$ through FoV), trial more DMs or increase the detection S/N (sensitivity) through more optimal dedispersion techniques. The threshold depends on the costs of forming and searching the beams, which increases with bandwidth. Also, the number of channels (hence bandwidth) required to reach this threshold reduces for steeper spectrum sources (see \autoref{tab:Contributing-channels}). 
For searches of populations where the event rate is increasing with frequency, such as directions of large scatter broadening, the first channel is at the highest frequency and further contributing channels come from lower frequencies.

Even if the search does not need the full band, recording it in a buffer is desirable to enable a more powerful analysis of detected transients with dedicated processing---this is especially important for relatively rare events. For flat spectrum sources, the S/N improves by a factor of $\sqrt{\Delta\nu}$. The extra spectral information is also a useful analysis tool. For example, the multipath propagation which causes scatter broadening would be more evident with a larger bandwidth, given the $\nu^{-4.4}$ relationship in \citet{CorLaz02}, allowing us to distinguish between the intrinsic pulse width and a pulse that has been scatter broadened.

\subsection{Application to Galactic populations\label{sec:Galactic}}

A search for a Galactic population of fast transients is dependent on observing direction. The first-order analysis of scatter broadening shows that the event rate varies as a function of observation direction and intrinsic pulse width. The dependence of event rate on frequency, described previously, still applies. For an intrinsic pulse width of 1~ms, low band dishes give a higher event rate than AA-low for directions on or near the Galactic plane where the scatter broadening is large. Although the effectiveness of AA-low is reduced at these directions, the increased luminosity of steep spectrum sources at lower frequencies somewhat counters this effect. Observations using the high band feed (1--2 GHz) on dishes are not modelled here, but in areas closer to the Galactic Centre, where scatter broadening is increased, it would achieve an event rate higher than the low band feed for spectrally shallow sources and sources with smaller intrinsic pulse widths.

The event rate per beam curve for populations near the Galactic plane (\autoref{fig:Event-rate-vs-freq}) is quite flat in comparison to extragalactic populations, especially for shallow spectrum sources. Because of this, additional processed bandwidth is more useful than for extragalactic populations, although it still cannot increase the event rate by more than the ideal case of $\Delta\nu^{3/4}$. For the threshold discussed in \autoref{sub:Event-rate-bandwidth}, more channels would contribute to the event rate.

Besides scatter broadening, another limit on the event rate occurs if the telescope is sensitive enough to observe the population to the edge of the Galaxy (i.e. the sources are luminous enough to be observable to the edge of the Galaxy). In that case, the event rate for a sensitivity limited volume (as used in this paper) is not valid, because the limit is instead imposed by the boundary of the Galaxy. \citet{Mac11} captures both of these effects by introducing a direction dependent factor $\delta$, such that
\begin{equation}
\mathcal{R}_{\nu}\propto \Omega_{\rm proc} S_{{\rm min}}^{-\frac{3}{2}+\delta}\quad {\rm events\,s}^{-1},\label{eq:Event-rate-delta}\end{equation}
where $0\le\delta\le 3/2$.
A key result in \citet{Mac11} is that FoV is more strongly preferred over sensitivity when scatter broadening or volume boundary limits increase $\delta$. Further modelling of the directional dependence of $\delta$ is shown in that paper. 

The total number of events detected in a transients survey is proportional to $\mathcal{R}_{\nu}t$, where $t$ is the total observation time. When $\delta$ is large, the contribution from $S_{\rm min}$ is lessened. In this case, the total number of events detected can be more effectively increased by forming multiple beams (if available), incoherently combining the array or subarray signals (gaining a much larger FoV that coherent combination) or simply spending more time observing the sky. A low cost commensal survey using the incoherently combined array achieves these, making it even more effective for populations near the Galactic plane.

To determine the number of events detectable for a Galactic population, simulations
must account for all frequency-dependent effects on $\mathcal{R}_{{\rm beam^{-1}}}$
as a function of sky direction. Simulations for LOFAR pulsar searches
\citep{LeeSta10} account for $T_{{\rm sky}}$ (which modifies telescope
sensitivity) as a function of direction and scatter broadening as a
function of direction and distance. For a population with 
a distribution of luminosities, the event rate is distance dependent. 
By numerically integrating along each line of sight, \citet{Mac11} accounts for sources at each distance step which can no longer be detected below a certain sensitivity due to scatter broadening. 
 Such an analysis is beyond the scope of this work, and requires assumptions to be made about the luminosity and spatial distribution of the objects, and their intrinsic pulse widths.

\subsection{Further work}
Maximising $\mathcal{R}_{\rm beam^{-1}}$ involves trade-offs between the receptor, signal combination mode, observing frequency and bandwidth for a given sky direction. This first-order analysis assumes $\mathcal{R}_{\rm cost^{-1}}\propto\mathcal{R}_{\rm beam^{-1}}$. The next logical step is to describe the cost of forming and searching beams, to maximise $\mathcal{R}_{\rm cost^{-1}}$. This problem is complex because beamforming and dedispersion processing costs are architecture specific.  However, modelling these costs will give a better understanding of the optimal bandwidth and frequency to use for a given survey strategy and sky direction.
Areas of work to further maximise $\mathcal{R}_{\rm cost^{-1}}$ are:
\begin{itemize}
\item Trade-offs between station size and number of stations, for fixed AA-low collecting area.
\item The S/N gain versus processing cost from using a more accurate incoherent dedispersion method.
\item The S/N gain versus processing cost of using coherent dedispersion. Incoherent dedispersion has a lower S/N ratio than coherent dedispersion, but requires significantly less processing. To maintain the same detection rate as coherent dedispersion, how many more beams need to be formed and searched?
\end{itemize}

\section{Conclusions\label{sec:Conclusions}}

Radio telescopes are inherently flexible, giving rise to a large design trade-off space, but any trade-offs must be made with the consideration of cost. This paper presents a new figure of merit,  event rate per beam ($\mathcal{R}_{\rm beam^{-1}}$), to measure the effectiveness of a survey strategy in detecting transient events in a volume of sky, while considering first-order costs---the ultimate goal being to calculate the event rate per unit cost $(\mathcal{R}_{\rm cost^{-1}})$. The results show the complexities in determining an optimal receptor,  signal combination mode and frequency range for a given array. They highlight some important points for searching for fast transients with SKA1 and more broadly with radio telescope arrays.

The per beam event rate enables a frequency independent analysis of the optimal signal combination mode and array configuration for a given receptor. This analysis is applied to SKA1, but regardless of the telescope used, incoherent combination always achieves a higher $\mathcal{R}_{\rm beam^{-1}}$, making it preferable to coherent combination, subarraying and fly's eye in most cases. This is due to the event rate for surveying a volume of sky increasing faster with FoV than with sensitivity. The advantage of incoherent combination increases with a lower filling factor or the combination of an array with more elements. 
The exception to this result can occur once the number of dish or station beams formed reaches its maximum, due to physical and processing constraints. In this case, forming more array beams can result in coherent combination achieving a higher total event rate than incoherent combination.
Fly's eye is unattractive because of the need to search many more beams, and its lack of localisation capability. However, if the search cost is low, incoherently combined three-element subarrays could usefully employ antennas unused by the primary (e.g. imaging) user observation. An example with SKA1 would be splitting 24 AA-low stations outside the core into 8 three-element subarrays, to achieve a total event rate similar to the incoherent combination of the total array. 

From \autoref{sec:Rate-modes} the per beam event rates for incoherent and coherent combination can be compared using \[
\mathcal{R}_{\rm coh\,beam^{-1}}  =  \frac{\eta^{3/2}N_{0}^{3/4}}{N_{\rm pix}}\mathcal{R}_{\rm inc\,beam^{-1}},\]
where $N_0$ is the number of stations or dishes used for incoherent combination, $\eta$ is the fraction of   $N_{0}$ used for coherent combination and $N_{\rm pix}$ is the number of coherently formed array beams required to fill the FoV of the station or dish beam. This reduces to
\[\mathcal{R}_{\rm coh\,beam^{-1}} = N_{0}^{-1/4}\,\mathcal{R}_{\rm inc\,beam^{-1}}\] for the ideal densely packed case, although the ratio is actually higher due to physical limitations.  If the $N_{0}$ antennas are divided into $N_{\rm sa}$ subarrays, \[\mathcal{R}_{\rm sa\,beam^{-1}} = N_{\rm sa}^{-3/4}\mathcal{R}_{\rm inc\,beam^{-1}}.\]

Dependence on observing frequency and bandwidth needs to be considered when making event rate calculations, as does the observation direction and expected spectrum of the source.
The results in this paper show that for commensal observing, where the direction of observation is determined by the primary telescope user, a fast transients pipeline must dynamically adjust the search strategy to achieve the highest event rate. For a targeted survey, the choice of receptor needs careful consideration.

For extragalactic searches with SKA1, the full available bandwidth does not need to be searched; it is not an optimal use of the processing system. The contribution to event rate from processing additional bandwidth decreases as the signal frequency increases. Beyond a threshold, the processing could be more effectively used to form and search more beams, trial more DMs or increase the detection S/N ratio. However, the full band is desirable for analysis of detected transients and for searches in directions where scatter broadening is such that the event rate is approximately constant with frequency.

\section{Recommendations for SKA1\label{sec:Recommendations}}
\begin{itemize}
\item \textbf{SKA processing needs to provide flexible search modes}

The preferred receptor and signal combination mode depends on direction, especially for Galactic populations. Incoherent and coherent combination modes are both effective with SKA1 and depend on the array filling factor, signal processing costs and the spectrum of the source population.

For SKA1, coherent combination of the AA-low core requires approximately five array beams to equal the event rate of a single beam of the incoherently combined total array. For low band dishes, this number is 157, making the coherent combination of dishes unattractive unless more than this many array beams can be formed.

SKA1 AA-low is effective for extragalactic transients searches (assuming dedispersion costs are not too high), especially for steep spectrum sources (see \autoref{fig:Event-rate-vs-freq}).  For flat spectrum sources, the advantage of aperture arrays disappears. Low band dishes give a higher event rate for directions near the Galactic plane, where scatter broadening is larger.

\item \textbf{The re-use of signal processing required for SKA imaging modes enables low cost fast transients searches}

Commensal surveys using the incoherent combination of dish signals or AA station beams are low cost options for searching for fast transients with SKA1. The incremental cost of implementing such a search is small because it uses beams formed for the primary user observation. A commensal survey using both receptors is a simple method to increase the total number of events detected, by increasing the total observation time. The survey effectiveness, compared to the coherent combination of the array, increases for populations closer to the Galactic plane.

Access to the dish and station beam data should be via a spigot with a well defined interface, to enable the implementation of flexible search modes. The processing for commensal surveys could be implemented internally or with user-provided processing as it becomes affordable or available.

\item \textbf{Requirements for fast transient searches with SKA1}

Until the processing costs are further explored, the base requirements for SKA1 are:
\begin{itemize}
\item Availability of incoherent and coherent combination modes for AA-low and low band dishes.
\item Processing for low cost commensal survey modes; or provision for access to the dish and AA station beam data via spigots.
\item Voltage (coherent) buffering capability of the full band; of order tens of seconds for dish frequencies and possibly minutes for lower frequencies.
\end{itemize}
For extragalactic searches, processing the full available bandwidth is not required. Bandwidths of 50--100 MHz are sufficient on the basis of the simplified investigation undertaken in this paper; a more detailed study of the trade-offs could be made. However, buffering the full band is desirable for analysis of detected transients.
\end{itemize}

\section*{Acknowledgements}
The authors thank R. D. Ekers, P. J. Hall, J-P. Macquart and S. Ord for discussions and suggestions. The International Centre for Radio Astronomy Research is a Joint Venture between Curtin University and The University of Western Australia, funded by the State Government of Western Australia and the Joint Venture partners. T. M. Colegate is a recipient of an Australian Postgraduate Award and a Curtin Research Scholarship.


\appendix
\section*{Appendices}
\numberwithin{equation}{section} 
\section{Signal combination techniques \label{sec:Signal-combination}}

The performance attributes of a radio telescope array depend on how the signals from the array elements are combined. This in turn affects the detection rate for fast transients. The array elements may be an antenna (such as a dish or dipole) or a phased group of antennas (stations). We term the single antenna primary beam or the phased station beam as the element beam, with FoV $\Omega_0$. The signals from these elements may then be combined incoherently or coherently as discussed later in this section.

\subsection{AA station beamforming\label{sub:Station-beamforming}}
An aperture array station will have thousands of individual antennas. To reduce the data rate from the station and the downstream signal processing load, the antennas will be phased into a station beam, as described in \citet{ZarFau10}. Multiple stations beams can be formed by applying the appropriate phase shift. The beamformer processing cost scales as
$N_{\rm a/st}N_{\rm b-0}\Delta\nu$ operations per second,
 where $N_{\rm a/st}$ is the number of antenna elements per station, $N_{\rm b-0}$ is the number of station beams formed and $\Delta\nu$ is the bandwidth \citep{Cor07}.

\subsection{Modes of beamforming for searching}

For antenna, phased array feed or station beams pointing at the same location on the sky, the signals detected may be combined incoherently or coherently. The following tables show how sensitivity, FoV,  beamformer processing cost and the number of data streams to be searched scale for different signal combination modes. The scaling equations assume that the polarisations are summed prior to searching and all elements are of equal diameter and sensitivity. 

\subsubsection{Incoherent combination}

\begin{figure}[h]
\begin{center}
\includegraphics[width=1\columnwidth]{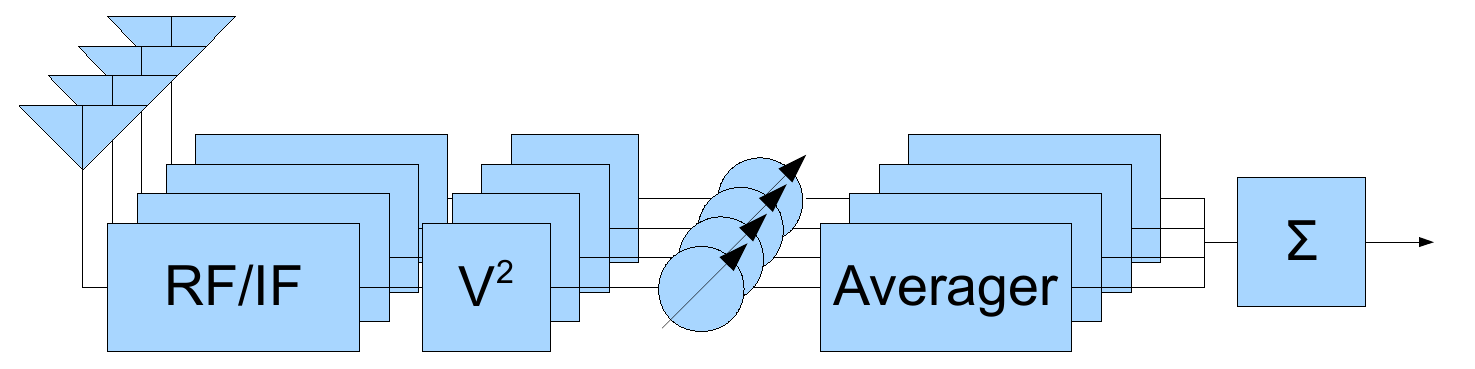}
\caption[Incoherent combination]{Incoherent combination. For aperture arrays, station beamforming takes
place prior to the $V^{2}$ block.\label{fig:Incoherent-combination}}
\end{center}
\end{figure}

Incoherent combination of the element signals requires the signal
from each element to be detected, a geometric delay applied and the
signals summed. 
\autoref{fig:Incoherent-combination} shows the steps to incoherently
combine signals and \autoref{tab:Incoherent-combination-attributes} shows some performance attributes. 

\begin{table}[h]
\begin{center}
\caption{Incoherent combination attributes \label{tab:Incoherent-combination-attributes}}
\begin{tabular}{>{\raggedright}p{0.3\columnwidth}>{\centering}p{0.2\columnwidth}>{\raggedright}p{0.4\columnwidth}} 
\hline
Attribute &  Scaling & Comment\tabularnewline
\hline
Sensitivity ($A_{\rm e}$) & $\sqrt{N_{0}}A_{\rm e0}$ & $N_{0}$ elements, each with effective area  $A_{\rm e0}$\tabularnewline
FoV ($\Omega_{\rm proc}$) & $N_{\rm b-0}\Omega_{0}$ & $N_{\rm b-0}$ beams formed per station, each with FoV $\Omega_{0}$\textsuperscript{a}\tabularnewline
Processing & $N_{0} N_{\rm b-0} \Delta\nu$ & \tabularnewline
Data streams & $N_{\rm b-0}$ &  \tabularnewline
\hline
\end{tabular}
\end{center}
\medskip
\textsuperscript{a} $\Omega_{0}= \frac{\pi}{4}\left(\frac{c\mathcal{K}_0}{\nu D_0}\right)^{2}$, where $\mathcal{K}_0$ is the dish illumination or station beam taper and $D_0$ is the diameter of the dish or station in metres.\\
\end{table}

As long as the appropriate geometric delay is applied to the signal
at each element, incoherently combined elements do not need to be
located close together. For the SKA, this means that while the core is
being used for low angular resolution experiments, the mid and long
baselines could be used for fast transient searches. Note that incoherent
combination cannot account for the geometric delays within the beam,
but away from the beam centre.

We assume that digitising, channelising and station beamforming are existing telescope functions and that these do not factor into the additional processing cost of incoherently combining the station beams.  Square-law detection involves squaring and summing the real and imaginary components of each channel of each beam from each station.  The averager integrates the power samples for each beam, and then equivalent beams from different elements are summed together, resulting in a total of $N_{\rm b-0}$ incoherently combined data streams. The integration of the power samples greatly reduces the data rate of the incoherent beams, although it comes at a cost of a lower time resolution.

\subsubsection{Independently pointed subarrays, incoherently combined}

A further increase in FoV can be achieved by pointing subarrays of elements in different directions and incoherently combining the signals from elements in the subarray.  \autoref{tab:Incoherent-subarray-attributes} shows performance attributes for this mode.  The processing cost for incoherently combining subarrays is approximately the same as it is for incoherently combining all stations because the same number of beams need to be square-law detected and summed across stations; the difference is that separate sums need to be maintained for each subarray.  The approximate beamforming operations cost is therefore independent of the number of incoherently combined subarrays ($N_{\rm sa}$).

\begin{table}[h]
\begin{center}
\caption{Incoherently combined subarray attributes.\label{tab:Incoherent-subarray-attributes}}
\begin{tabular}{>{\raggedright}p{0.3\columnwidth}>{\centering}p{0.2\columnwidth}>{\raggedright}p{0.4\columnwidth}} 
\hline
Attribute &  Scaling & Comment\tabularnewline
\hline
Sensitivity ($A_{\rm e}$) & $\sqrt{N_{\rm 0/sa}}A_{\rm e0}$ & $N_{\rm 0/sa}$ elements per subarray\tabularnewline
FoV ($\Omega_{\rm proc}$) & $N_{\rm sa} N_{\rm b-0}\Omega_{0}$  & $N_{\rm sa}$ independently pointed subarrays \tabularnewline
Processing &  $N_{0}N_{\rm b-0} \Delta\nu $ & \tabularnewline
Data streams & $N_{\rm sa}N_{\rm b-0}$ & \tabularnewline
\hline
\end{tabular}
\medskip\\
\end{center}
\end{table}

The smallest subarray size is one element, meaning each element is pointing to a unique patch
of sky --- this is termed fly's eye. In reality, the minimum number
of elements in a subarray would be 3, to allow for a triggered buffer
to be used to localise any detected signal.

\subsubsection{Coherent combination---array beamforming}

\begin{figure}[h]
\begin{center}
\includegraphics[width=1\columnwidth]{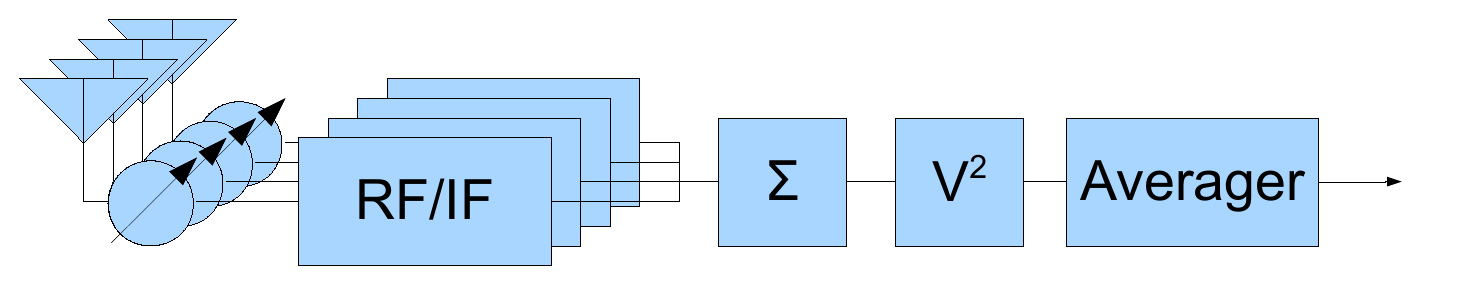}

\caption[Coherent combination.]{Coherent combination. For aperture arrays, station beamforming takes
place prior to the $\sum$ block.\label{fig:Coherent-combination}}
\end{center}
\end{figure}

For coherent combination, each array beam is formed by the weighted sum of $N_{0}$ element beams pointing
in the same direction. Geometric delays are applied to the signals from the elements, which are then summed and detected. Like AA stations, multiple beams can be formed.  \autoref{fig:Coherent-combination}
shows the steps for coherent array beamforming, and \autoref{tab:Coherent-combination} shows the performance attributes. \citet{Cor07} discusses array beamforming in the context of the SKA.

\begin{table}[h]
\begin{center}
\caption{Coherent combination attributes.\label{tab:Coherent-combination}}
\begin{tabular}{>{\raggedright}p{0.3\columnwidth}>{\centering}p{0.2\columnwidth}>{\raggedright}p{0.4\columnwidth}} 
\hline
Attribute &  Scaling & Comment\tabularnewline
\hline
Sensitivity ($A_{\rm e}$) & $N_{0}A_{\rm e0}$ &  Coherent sum of $N_{0}$ elements\tabularnewline
FoV ($\Omega_{\rm proc}$) & $N_{\rm b-arr}\Omega_{\rm arr}$\textsuperscript{a} & $N_{\rm b-arr}$ array beams formed \tabularnewline
Processing & $N_{0}N_{\rm b-arr}\Delta\nu$ & \tabularnewline
Data streams & $N_{\rm b-arr}$ & \tabularnewline
Data streams to pixelise full FoV &  $N_{\rm b-0}N_{\rm pix}$ & $N_{\rm pix}$ pixels to fill $\Omega_{0}$ with array beams  (see \autoref{eq:N_pix})\tabularnewline
\hline
\end{tabular}
\end{center}
\medskip
\textsuperscript{a}$\Omega_{\rm arr}=\frac{\pi}{4}\left(\frac{c\mathcal{K}_{\rm arr}}{\nu D_{\rm arr}}\right)^{2}$, where  $\mathcal{K}_{\rm arr}$ is the array beam taper and  $D_{\rm arr}$  is the longest baseline in the array.
\end{table}

\subsubsection{Correlation beamforming---{}`fast imaging'}

An alternative method to pixelise the full FoV is to create
images from the correlator at a high time resolution. The highest
time resolution available is specified by the correlation integration
period, which must be short enough to prevent smearing. \citet{Cor07} 
compares the computational cost of this method with the coherent
sum, and finds that the cost-effectiveness depends on configuration
and system temperature. However the major advantage is that the correlator
hardware can be re-used. 
It should be noted that this means that the data must be able to be
dumped from the correlator at these high rates which may place additional
requirements on the correlator architecture. There is also a computational cost for gridding and imaging at high time resolution.

\begin{table}[h]
\begin{center}
\caption{Correlation beamforming attributes.}
\begin{tabular}{>{\raggedright}p{0.3\columnwidth}>{\centering}p{0.2\columnwidth}>{\raggedright}p{0.4\columnwidth}} 
\hline
Attribute &  Scaling & Comment\tabularnewline
\hline
Sensitivity ($A_{\rm e}$) & $N_{0}A_{\rm e0}$ & For large $N_{0}$. \tabularnewline
FoV ($\Omega_{\rm proc}$) & $N_{\rm b-0}\Omega_{0}$ & \tabularnewline
Processing & $1$ & Use existing correlator hardware\tabularnewline
Data streams & $N_{\rm b-0}N_{\rm pix}$ & \tabularnewline 
\hline
\end{tabular}
\end{center}
\medskip
\end{table}

\section{Event rate for a broadened pulse in a volume of sky\label{sec:Rate-derivation}}
The detected event rate is calculated for observable space. We assume
extragalactic sources of intrinsic luminosity $\mathcal{L}_{\rm i}\,{\rm Jy\, pc^{2}}$
are homogeneously distributed in a sphere of volume $V$ with a nominal
event rate density of $\rho\, {\rm events\,s^{-1}\,pc^{-3}}$. The event rate for this sphere is given by $\rho V$:\begin{equation}\mathcal{R}_{\rm sphere} = \rho\frac{4\pi}{3}D^{3}\quad {\rm events\, s^{-1}},\end{equation}
\-where $D$ is the radius of the sphere in pc. From \citet{DenCor09},
a source can be detected out to a maximum distance\begin{equation}
D_{\rm max}=\sqrt{\frac{\mathcal{L}_{\rm i}}{4\pi S_{\rm i}}},\end{equation}
where $S_{\rm i}$ is the intrinsic flux density of the source.

The intrinsic flux density of a pulse differs from the observed flux density due to pulse broadening (or smearing) effects of the interstellar medium and of the detection system itself.  However, if no energy is lost due to attenuation, then pulse ``area'' is conserved such that $S_{\rm i}W_{\rm i}=SW$, 
where $W_{\rm i}$ is the intrinsic width of the pulse, $S$  is the observed flux density and $W$ is the width of the broadened pulse.  If the post-detection integration time $\tau$ is equal to $W$, a telescope with minimum detectable flux density $S_{{\rm min},\tau=W}$ can detect the broadened pulse to a maximum distance
\begin{equation}
D_{\rm max}=\sqrt{\frac{W_{\rm i}\mathcal{L}_{\rm i}}{4\pi WS_{{\rm min},\tau=W}}}.
\end{equation}
A more general relationship which uses an integration time of $\tau=W_{\rm i}$ is
\begin{equation}
D_{\rm max}=\left({\frac{W_{\rm i}}{W}}\right)^\frac{1}{4}\left(\frac{\mathcal{L}_{\rm i}}{4\pi S_{{\rm min}}}\right)^\frac{1}{2}.
\end{equation}

Pulse broadening factors are discussed further in \autoref{sec:Broadening}. 

The nominal extragalactic population is observable out to $D_{\rm max}$
for the fraction of the sky observed ($\Omega_{\rm proc}/\Omega_{\rm sky}$)
and the detected event rate is \begin{align}\begin{split}
\mathcal{R}_{ \rm \nu} & = \frac{4\pi}{3}\rho\frac{\Omega_{\rm proc}}{\Omega_{\rm sky}}D_{\rm max}^{3}\\
 & = \frac{1}{3}\rho\Omega_{\rm proc}\left(\frac{W_{\rm i}}{W}\right)^\frac{3}{4}\left(\frac{\mathcal{L}_{\rm i}}{4\pi S_{{\rm min}}}\right)^{\frac{3}{2}}.\end{split}\end{align}
$\Omega_{\rm proc}$, $S_{{\rm min}}$ and $W$ are functions of how the signals are combined and processed by the telescope system.  $S_{{\rm min}}=\sigma\Delta S_{\rm rms}$, where
$\sigma$ is the required S/N ratio and $\Delta S_{\rm rms}$ is the rms variation
in flux density for a randomly polarised source, obtained by applying
the telescope gain to the radiometer equation:\begin{equation}
S_{{\rm min}}=\sigma\Delta S_{\rm rms}=\frac{\sigma2k_{\rm B}T_{\rm sys}}{A_{\rm e}\sqrt{N_{\rm pol}\Delta\nu\tau}},\end{equation}
where $T_{\rm sys}$ is the system temperature, $A_{\rm e}$ is the total
effective area used in the survey, $\Delta\nu$ is the processed bandwidth and
$\tau$ is the post-detection integration time, which also defines the time resolution. 

\section{Pulse broadening and correction (dedispersion)}\label{sec:Broadening}
\citet{CorMcl03} model the broadening of a delta function pulse due to propagation through the interstellar medium and signal processing response times using the following approximation:\begin{equation}
\Delta t = \sqrt{\Delta t_{\rm DM}^2 + \Delta t_{\rm \delta DM}^2 + \Delta t_{\Delta \nu}^2 + \tau_d^2},\end{equation}
where $\Delta t_{\rm DM}$ represents the broadening due to dispersion smearing; $\Delta t_{\rm \delta DM}$ is due to the error in the DM, $\delta DM$, used by the system's dedispersion signal processing; $\Delta t_{\Delta \nu}$ is the system's filter response time; and $\tau_d$ is due to the multipath propagation effects of the medium.  Expanding on this, the broadening of a pulse of intrinsic width $W_{\rm i}$ can be modelled as\begin{equation}
W = \sqrt{W_i^2 + \Delta t_{\rm DM}^2 + \Delta t_{\rm \delta DM}^2 + \Delta t_{\Delta \nu}^2 + \tau_d^2}.\end{equation}

The filter response time, $\Delta t_{\Delta \nu}$, is approximately equal to ${\Delta\nu}^{-1}$, where $\Delta\nu$ is the bandwidth of the filtered signal.  It is important to note that for a fully coherent transient detection system where the element beams are coherently combined, coherently dedispersed and searched, $\Delta\nu$ represents the full signal bandwidth; whereas for an incoherent transient detection system in which the signal is channelised, detected and searched, $\Delta\nu$ represents the much smaller channel bandwidth.  Consequently, the filter response component of pulse broadening, $\Delta t_{\Delta \nu}$, is significantly higher for incoherent transient detection systems.

Furthermore, while coherent dedispersion techniques can completely correct for dispersion smearing (given that the DM is known), incoherent dedispersion techniques can only correct for dispersion between the filter-bank channels; they cannot correct for dispersion within the channels.  Intra-channel dispersion smearing can be reduced by choosing smaller channel bandwidths, but at the expense of larger filter response times. The optimum channel bandwidth for incoherent dedispersion occurs where the dispersion smearing within each channel equals the filter response time  \citep{HanRic75, CorMcl03}.  This leads to a minimum pulse width after incoherent dedispersion of\begin{equation}
W_{\rm inc} = \sqrt{W_i^2 + 2(\Delta t_{\rm DMmin})^2 + \Delta t_{\rm \delta DM}^2 + \tau_d^2},\end{equation}
where $\Delta t_{\rm DMmin} = \sqrt{8.3\times10^{15}DM\nu^{-3}}$.  It should be noted that this optimum cannot be realised for all DMs, because it expects the channel bandwidth to be a function of the DM.

For coherent dedispersion the $\Delta t_{\rm DM}$ term is completely removed:\begin{equation}
W_{\rm coh} = \sqrt{W_i^2 + \Delta t_{\Delta\nu}^2 + \Delta t_{\rm \delta DM}^2 + \tau_d^2},\end{equation}
where (as noted above) $\Delta\nu$ is the full processed bandwidth and $\Delta t_{\Delta\nu} \approx \Delta\nu^{-1}$.

\section{Frequency dependence of low frequency aperture arrays\label{sub:AA-low-Appendix}}

\begin{figure*}[t]
\begin{center}
\includegraphics[height=0.26\textheight]{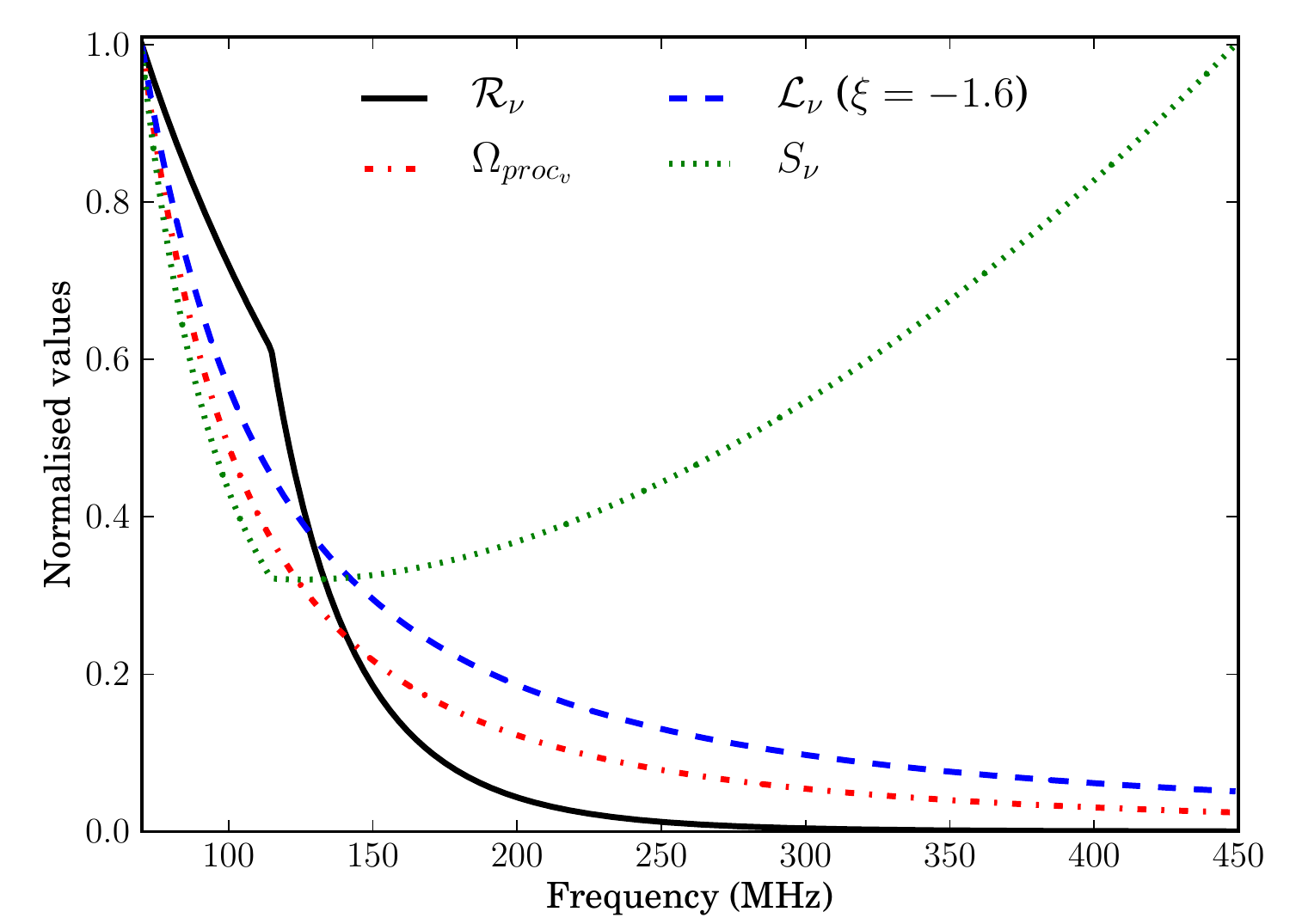}\includegraphics[height=0.26\textheight]{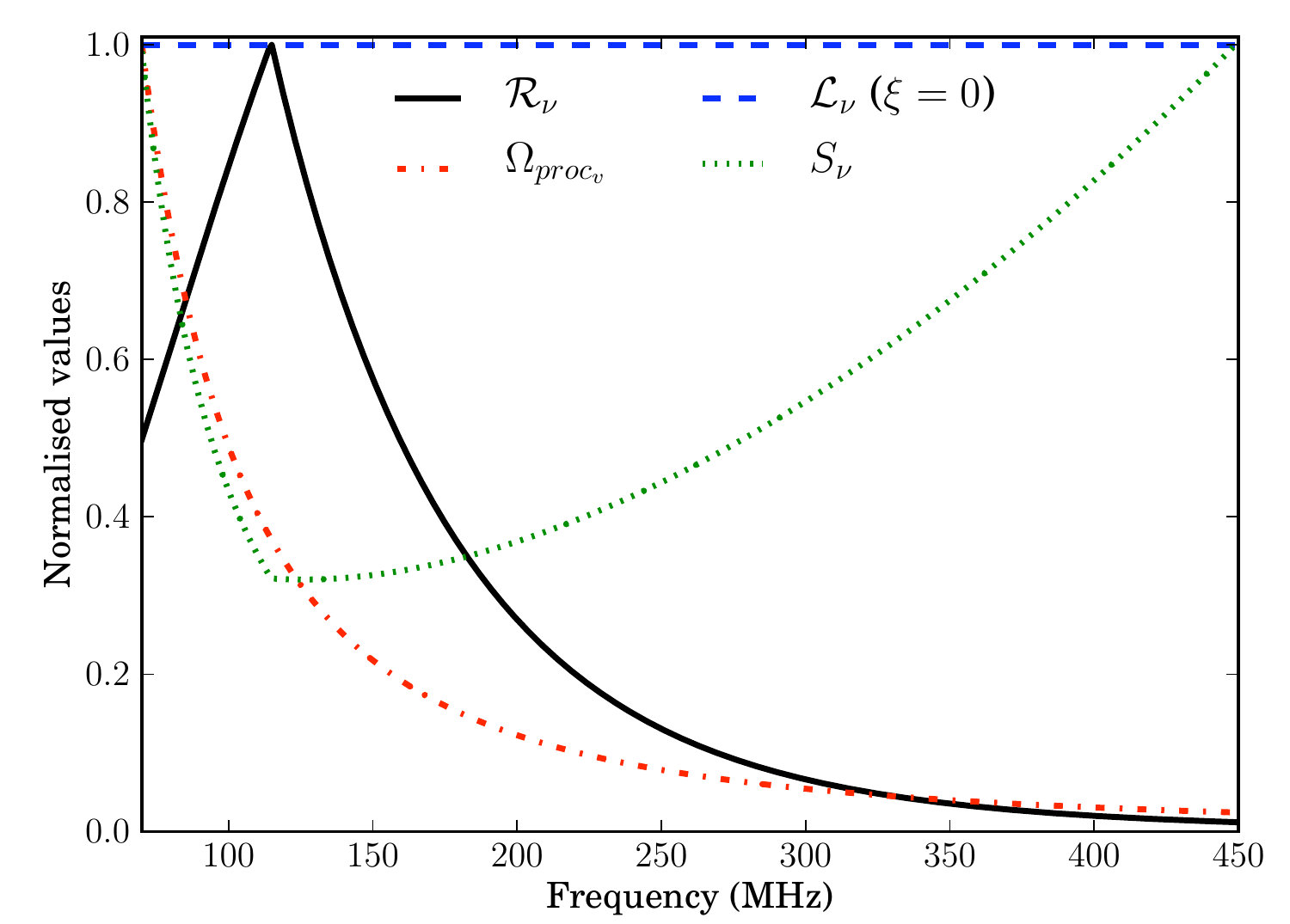}
\caption{Event rate $\mathcal{R}_{\nu}$ for $\xi=-1.6$ (left) and $\xi=0$ (right), and breakdown of the frequency-dependent components comprising $\mathcal{R}_{\nu}$, normalised to the maximum value of each.\label{fig:RateVsFreq-AA}}
\end{center}
\end{figure*}

\begin{figure*}[t]
\begin{center}
\includegraphics[height=0.26\textheight]{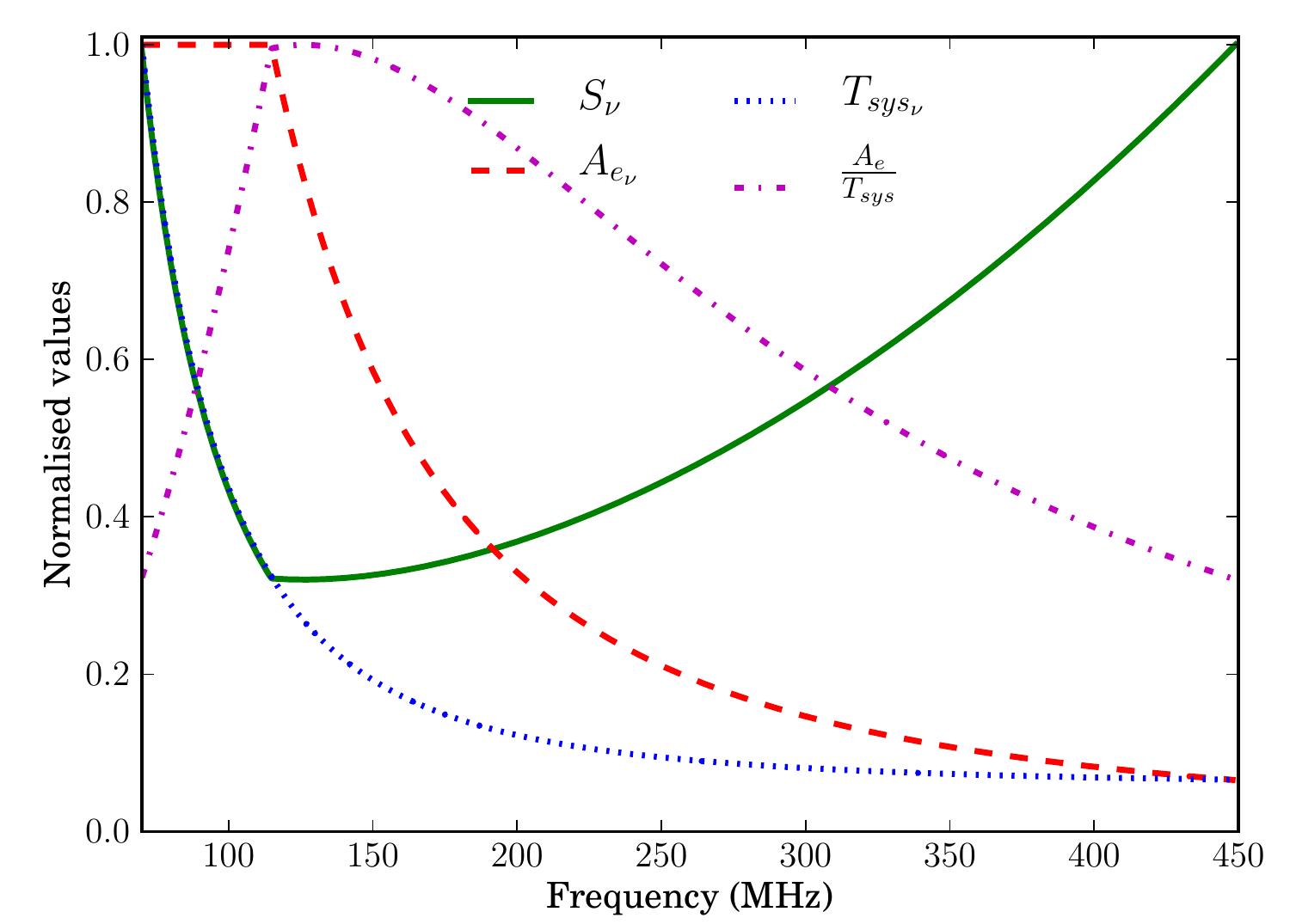}\caption{$S_{{\rm min}_{\nu}}$ and breakdown of the frequency-dependent components
comprising $S_{{\rm min}_{\nu}}$,  normalised to the maximum value of each.\label{fig:RateVsFreq-AA-Smin}}
\end{center}
\end{figure*}

For illustrative purposes, we plot the event rate $\mathcal{R}_{\nu}$ and a breakdown of its frequency-dependent components. \autoref{fig:RateVsFreq-AA}
shows $\mathcal{R}_{\nu}$ for source luminosities with spectral indices $\xi=-1.6$ and 0, over a frequency range 70--450~MHz, at 1~MHz steps with processed bandwidth $\Delta\nu=1\, {\rm MHz}$ and  normalised to $\mathcal{R}_{\nu}=1$ at 70~MHz.  The
slope is steep: at 160 MHz, the event rate is 10\% of the event 
rate at 70 MHz. At 450 MHz, the event rate is 0.016\% of
the 70 MHz event rate. A further breakdown of $S_{{\rm min}_{\nu}}$ is shown in \autoref{fig:RateVsFreq-AA-Smin}.

\section{Event rate as a function of frequency\label{ap:App-Event-Rate}}
The event rate is determined numerically by calculating the S/N ratio (SNR) for each
of $N_{\rm ch}$ frequency channels of width $\Delta\nu_{\rm ch}$, weighting
it by $w$, the amount of sky seen with that SNR and then taking the
root of the sum of the squares:\begin{equation}
\mathcal{R}_{\Delta v}=\sqrt{\sum_{i}^{N_{\rm ch}}\left(\mathcal{C}w_{i}SNR_{i}\right)^{2}},\end{equation}
where $\mathcal{C}$ is a constant. Given $SNR=\frac{\mathcal{L}}{4\pi S}$,\begin{equation}
\mathcal{R}_{\nu}^{2/3} = \left(\frac{1}{3}\rho\Omega_{\rm proc_{\nu}}\right)^{\frac{2}{3}}\frac{\mathcal{L}_{\nu}}{4\pi S_{{\rm min}_{\nu}}}.\end{equation}
Setting $w=\Omega_{\rm proc}^{2/3}$ and $\mathcal{C}=(\frac{1}{3}\rho)^{2/3}$
so $\mathcal{R}_{\nu}^{2/3}=\mathcal{C}\times w\times SNR,$
we can say that 
 \begin{align}\begin{split}
\mathcal{R}_{\Delta v} & =\left(\sum_{i}^{N_{\rm ch}}\mathcal{R}_{\nu,i}^{4/3}\right)^{\frac{3}{4}}\\
& =\frac{1}{3}\rho\left(\frac{\mathcal{L}_{0}}{\nu_{0}^{\xi}}\right)^{\frac{3}{2}}\left(\sum_{i}^{N_{\rm ch}}\left(\frac{\Omega_{{\rm proc},i}^{2/3}\nu_{i}^{\xi}}{S_{{\rm min},i}}\right)^{2}\right)^{\frac{3}{4}},\end{split}\end{align} 
for a processed bandwidth of $\Delta\nu=N_{\rm ch}\Delta\nu_{\rm ch}$. Where there is no frequency dependence, the event rate for total bandwidth
$\Delta\nu$ and unit bandwidth $\Delta\nu_{\rm ch}=1$ becomes\begin{equation}
\mathcal{R}_{\Delta\nu}=\frac{1}{3}\rho\Delta\nu^{3/4}\Omega_{\rm proc}\left(\frac{\mathcal{L}}{S_{\rm min}}\right)^{\frac{3}{2}}\end{equation}
as expected.


\begin{thebibliography}{26}

\bibitem[{{Burke-Spolaor} \& {Bailes}(2010)}]{BurBai10}
{Burke-Spolaor}, S., \& {Bailes}, M. 2010, MNRAS, 402, 855

\bibitem[{Chippendale {et~al.}(2007)Chippendale, Colegate, \&
  O'Sullivan}]{ChiCol07}
Chippendale, A.~P., Colegate, T.~M., \& O'Sullivan, J.~D. 2007, SKAcost: a tool
  for {SKA} cost and performance estimation, SKA Memo 92

\bibitem[{{Clarke} {et~al.}(2011){Clarke}, {D'Addario}, {Navarro}, {Cheng}, \&
  {Trinh}}]{ClaDAd11}
{Clarke}, N.~L., {D'Addario}, L., {Navarro}, R., {Cheng}, T.-H., \& {Trinh}, J.
  2011, An Architecture for Incoherent Dedispersion, CRAFT Memo 6

\bibitem[{Cordes(2009)}]{Cor07}
Cordes, J.~M. 2009, The {SKA} as a Radio Synoptic Survey Telescope: Widefield
  Surveys for Transients, Pulsars and {ETI}, 14th edn., SKA Memo 97

\bibitem[{{Cordes} \& {Lazio}(2002)}]{CorLaz02}
{Cordes}, J.~M., \& {Lazio}, T.~J.~W. 2002, arXiv:astro-ph/0207156

\bibitem[{Cordes \& McLaughlin(2003)}]{CorMcl03}
Cordes, J.~M., \& McLaughlin, M.~A. 2003, ApJ, 596, 1142

\bibitem[{{Cordes} {et~al.}(2004){Cordes}, {Lazio}, \&
  {McLaughlin}}]{CorLaz04} 
  Cordes, J.~M., Lazio,  T.~J.~W. \& McLaughlin, M.~A., 2004, NewAR, 48, 1459

\bibitem[{{Cordes} {et~al.}(2006){Cordes}, {Freire}, {Lorimer}, {Camilo},
  {Champion}, {Nice}, {Ramachandran}, {Hessels}, {Vlemmings}, {van Leeuwen},
  {Ransom}, {Bhat}, {Arzoumanian}, {McLaughlin}, {Kaspi}, {Kasian}, {Deneva},
  {Reid}, {Chatterjee}, {Han}, {Backer}, {Stairs}, {Deshpande}, \&
  {Faucher-Gigu{\`e}re}}]{CorFre06}
{Cordes}, J.~M., {et~al.} 2006, ApJ, 637, 446

\bibitem[{{D'Addario}(2010)}]{DAd10}
{D'Addario}, L. 2010, {ASKAP} Surveys for Transients: Which Observing Mode is
  Best?, SKA Memo 123

\bibitem[{{Deneva} {et~al.}(2009){Deneva}, {Cordes}, {McLaughlin}, {Nice},
  {Lorimer}, {Crawford}, {Bhat}, {Camilo}, {Champion}, {Freire}, {Edel},
  {Kondratiev}, {Hessels}, {Jenet}, {Kasian}, {Kaspi}, {Kramer}, {Lazarus},
  {Ransom}, {Stairs}, {Stappers}, {van Leeuwen}, {Brazier}, {Venkataraman},
  {Zollweg}, \& {Bogdanov}}]{DenCor09}
{Deneva}, J.~S., {et~al.} 2009, ApJ, 703, 2259

\bibitem[{{Dewdney} {et~al.}(2010){Dewdney}, {bij de Vaate}, {Cloete}, {Gunst},
  {Hall}, {McCool}, {Roddis}, \& {Turner}}]{Dewbij10}
{Dewdney}, P.~E., {bij de Vaate}, J.-G., {Cloete}, K., {Gunst}, A.~W., {Hall},
  D., {McCool}, R., {Roddis}, N., \& {Turner}, W. 2010, {SKA Phase 1}:
  Preliminary System Description, SKA Memo 130

\bibitem[{{Faulkner} {et~al.}(2010){Faulkner} {et~al.}}]{FauAle10}
{Faulkner}, A.~J., {et~al.} 2010, Aperture Arrays for the {SKA}: the {SKADS}
  White Paper, SKA Memo 122

\bibitem[{{Garrett} {et~al.}(2010){Garrett}, {Cordes}, {Deboer}, {Jonas},
  {Rawlings}, \& {Schilizzi}}]{GarCor10}
{Garrett}, M.~A., {Cordes}, J.~M., {Deboer}, D.~R., {Jonas}, J.~L., {Rawlings},
  S., \& {Schilizzi}, R.~T. 2010, A Concept Design for {SKA Phase 1 (SKA1)},
  SKA Memo 125

\bibitem[{Graham {et~al.}(1998)Graham, Lubachevsky, Nurmela, \&
  Ostergard}]{GraLub98}
Graham, R.~L., Lubachevsky, B.~D., Nurmela, K.~J., \& Ostergard, P.~R.~J. 1998,
  Disc. Math., 181, 139

\bibitem[{{Hall} {et~al.}(2008){Hall}, {Schilizzi}, {Dewdney}, \& {Lazio}}]{HalSch08}
  {Hall}, P.~J., {Schilizzi}, R.~T., {Dewdney}, P.~E. \& {Lazio}, T.~J.~W. 2008, The Radio Science Bulletin, 326

\bibitem[{{Hankins} \& {Rickett}(1975)}]{HanRic75}
{Hankins}, T.~H., \& {Rickett}, B.~J. 1975, in Methods in Computational
  Physics, ed. {B.~Alder, S.~Fernbach, \& M.~Rotenberg}, Vol.~14, 55--129

\bibitem[{{Hessels} {et~al.}(2009){Hessels}, {Stappers}, {van Leeuwen}, \&
  {Transients Key Science Project}}]{HesSta09}
{Hessels}, J.~W.~T., {Stappers}, B.~W., {van Leeuwen}, J., \& {Transients Key
  Science Project}, L. 2009, in The Low-Frequency Radio Universe, ed. D.~J.
  Saikia, D.~Green, Y.~Gupta, \& T.~Venturi

\bibitem[{{Keane} {et~al.}(2011){Keane}, {Kramer}, {Lyne}, {Stappers}, \&
  {McLaughlin}}]{KeaKra11}
{Keane}, E.~F., {Kramer}, M., {Lyne}, A.~G., {Stappers}, B.~W., \& {McLaughlin}, M.~A. 2011, MNRAS,
  in press

\bibitem[{{Keith} {et~al.}(2010){Keith}, {Jameson}, {van Straten}, {Bailes},
  {Johnston}, {Kramer}, {Possenti}, {Bates}, {Bhat}, {Burgay}, {Burke-Spolaor},
  {D'Amico}, {Levin}, {McMahon}, {Milia}, \& {Stappers}}]{KeiJam10}
{Keith}, M.~J., {et~al.} 2010, MNRAS, 409, 619

\bibitem[{{Kijak} {et~al.}(2011){Kijak}, {Lewandowski}, {Maron}, {Gupta}, \& {Jessner}}]{KijLew11}
  {Kijak}, J., {Lewandowski}, W., {Maron}, O., {Gupta}, Y. \& {Jessner}, A. 2011, A\&A, 531, A16
  
\bibitem[{Lorimer {et~al.}(2007)Lorimer, Bailes, McLaughlin, Narkevic, \&
  Crawford}]{LorBai07}
Lorimer, D.~R., Bailes, M., McLaughlin, M.~A., Narkevic, D.~J., \& Crawford, F.
  2007, Sci, 318, 777

\bibitem[{{Lorimer} {et~al.}(1995){Lorimer}, {Yates}, {Lyne}, \&
  {Gould}}]{LorYat95}
{Lorimer}, D.~R., {Yates}, J.~A., {Lyne}, A.~G., \& {Gould}, D.~M. 1995, MNRAS,
  273, 411

\bibitem[{{Macquart} {et~al.}(2010a){Macquart}, {Hall}, \&
  {Clarke}}]{MacHal10}
{Macquart}, J.-P., {Hall}, P.~J., \& {Clarke}, N. 2010a, in
  {``International SKA Forum 2010 Science Meeting'' PoS(ISKAF2010)03}, Assen, the
  Netherlands
  
\bibitem[{{Macquart} {et~al.}(2010b){Macquart}, {Bailes}, {Bhat},
  {Bower}, {Bunton}, {Chatterjee}, {Colegate}, {Cordes}, {D'Addario}, {Deller},
  {Dodson}, {Fender}, {Haines}, {Halll}, {Harris}, {Hotan}, {Jonston}, {Jones},
  {Keith}, {Koay}, {Lazio}, {Majid}, {Murphy}, {Navarro}, {Phillips}, {Quinn},
  {Preston}, {Stansby}, {Stairs}, {Stappers}, {Staveley-Smith}, {Tingay},
  {Thompson}, {van Straten}, {Wagstaff}, {Warren}, {Wayth}, {Wen}, \& {CRAFT
  Collaboration}}]{MacBai10}
{Macquart}, J.-P., {et~al.} 2010b, PASA, 27, 272

\bibitem[{{Macquart}(2011)}]{Mac11}
{Macquart}, J.-P. 2011, ApJ, 734, 20

\bibitem[{{Malofeev} {et~al.}(1994){Malofeev}, {Gil}, {Jessner}, {Malov},
  {Seiradakis}, {Sieber}, \& {Wielebinski}}]{MalGil94}
  {Malofeev}, V.~M., {Gil}, J.~A., {Jessner}, A., {Malov}, I.~F., {Seiradakis}, J.~H., {Sieber}, W., \& {Wielebinski}, R. 1994, A\&A, 285, 201
  
\bibitem[{{Siemion} {et~al.}(2011){Siemion}, {Bower}, {Dexter}, {Foster},
  {Mallard}, {McMahon}, {Wagner}, {Werthimer}, \& {Allen Telescope Array
  Team}}]{SieBow11}
{Siemion}, A., {et~al.} 2011, in BAAS, Vol.~43, American Astronomical Society
  Meeting Abstracts \#217, 240.06

\bibitem[{Smits {et~al.}(2009)Smits, Kramer, Stappers, Lorimer, Cordes, \&
  Faulkner}]{SmiKra09}
Smits, R., Kramer, M., Stappers, B., Lorimer, D.~R., Cordes, J., \& Faulkner,
  A. 2009, A\&A, 493, 1161

\bibitem[{{Stappers} {et~al.}(2011){Stappers}, {Hessels}, {Alexov}, {Anderson},
  {Coenen}, {Hassall}, {Karastergiou}, {Kondratiev}, {Kramer}, {van Leeuwen},
  {Mol}, {Noutsos}, {Romein}, {Weltevrede}, {Fender}, {Wijers}, {B{\"a}hren},
  {Bell}, {Broderick}, {Daw}, {Dhillon}, {Eisl{\"o}ffel}, {Falcke},
  {Griessmeier}, {Law}, {Markoff}, {Miller-Jones}, {Scheers}, {Spreeuw},
  {Swinbank}, {Ter Veen}, {Wise}, {Wucknitz}, {Zarka}, {Anderson}, {Asgekar},
  {Avruch}, {Beck}, {Bennema}, {Bentum}, {Best}, {Bregman}, {Brentjens}, {van
  de Brink}, {Broekema}, {Brouw}, {Br{\"u}ggen}, {de Bruyn}, {Butcher},
  {Ciardi}, {Conway}, {Dettmar}, {van Duin}, {van Enst}, {Garrett}, {Gerbers},
  {Grit}, {Gunst}, {van Haarlem}, {Hamaker}, {Heald}, {Hoeft}, {Holties},
  {Horneffer}, {Koopmans}, {Kuper}, {Loose}, {Maat}, {McKay-Bukowski},
  {McKean}, {Miley}, {Morganti}, {Nijboer}, {Noordam}, {Norden}, {Olofsson},
  {Pandey-Pommier}, {Polatidis}, {Reich}, {R{\"o}ttgering}, {Schoenmakers},
  {Sluman}, {Smirnov}, {Steinmetz}, {Sterks}, {Tagger}, {Tang}, {Vermeulen},
  {Vermaas}, {Vogt}, {de Vos}, {Wijnholds}, {Yatawatta}, \&
  {Zensus}}]{StaHes11}
{Stappers}, B.~W., {et~al.} 2011, A\&A, 530, A80

\bibitem[{{van Leeuwen} \& {Stappers}(2010)}]{LeeSta10}
{van Leeuwen}, J., \& {Stappers}, B.~W. 2010, A\&A, 509, A7

\bibitem[{{Wayth} {et~al.}(2011){Wayth}, {Brisken}, {Deller}, {Majid},
  {Thompson}, {Tingay}, \& {Wagstaff}}]{WayBri11}
{Wayth}, R.~B., {Brisken}, W.~F., {Deller}, A.~T., {Majid}, W.~A., {Thompson},
  D.~R., {Tingay}, S.~J., \& {Wagstaff}, K.~L. 2011, ApJ, in press

\bibitem[{{Wilkinson} {et~al.}(2004){Wilkinson}, {Kellermann}, {Ekers},
  {Cordes}, \& {Lazio}}]{WilKel04}
{Wilkinson}, P.~N., {Kellermann}, K.~I., {Ekers}, R.~D., {Cordes}, J.~M., \&
  {Lazio}, T.~J.~W. 2004, NewAR, 48, 1551

\bibitem[{{Zarb Adami} {et~al.}(2010){Zarb Adami}, {Faulkner}, {bij de Vaate},
  {Kant}, \& {Pickard}}]{ZarFau10}
{Zarb Adami}, K., {Faulkner}, A., {bij de Vaate}, J.~G., {Kant}, G.~W., \&
  {Pickard}, P. 2010, in Phased Array Systems and Technology (ARRAY), 2010 IEEE
  International Symposium on, 883--890
\end{thebibliography}
\end{document}